\documentclass[%
onecolumn,%
oneside,%
floats,%
aps,%
prd,%
nobibnotes,%
nofootinbib,%
amsmath,%
amssymb,%
amsfonts,%
amscd,%
superscriptaddress,%
eqsecnum%
]{revtex4-1}
\usepackage[utf8]{inputenc}
\usepackage{graphicx,array,dcolumn}
\usepackage[paperwidth=210mm,paperheight=297mm,centering,hmargin=1.8cm,vmargin=2.5cm]{geometry}
\usepackage{url}

\def\beq#1{\begin{equation}\label{#1}}
\def\eeq{\end{equation}}
\def\beqa#1{\begin{eqnarray}\label{#1}}
\def\eeqa{\end{eqnarray}}

\def\comment#1{\relax}

\newcommand{\be}{\begin{equation}}
\newcommand{\ee}{\end{equation}}

\def\be{\begin{eqnarray}}
\def\ee{\end{eqnarray}}

\def\-g{\sqrt{-g}}

\renewcommand\rho{\varrho}

\begin{document}

\title{
Latest News from HST and JWST: Possible Explanation of the Impossibles. 
}

\author{A.D. Dolgov}
\email{dolgov@fe.infn.it}
\affiliation{Novosibirsk State University, Novosibirsk, 630090, Russia}
\affiliation{Bogolyubov Laboratory of Theoretical Physics, JINR, Dubna, 141980, Russia}

\begin{abstract}

Recent data released by James Webb Space Telescope (JWST) and, somewhat earlier, the data presented by  
Hubble Space Telescope (HST) are commonly understood as a strong indication for breaking of the canonical 
$\Lambda$CDM cosmology.  It is argued in the present work that massive primordial black holes (PBH) could  
seed  galaxy and quasar formation in the very young universe as it has been conjectured in our paper of 1993 
and resolve the tension induced by the JWST and the HST data with the standard cosmology. This point of view 
is  rediscovered in several recent works. The proposed mechanism of PBH formation leads to the log-normal 
mass spectrum of PBHs, predicts noticeable antimatter population of our Galaxy, Milky Way. We also predict an
existence of central black holes of intermediate mass in dwarf galaxies and globular clusters, recently observed. 
All these predictions are in excellent agreement with astronomical observations in the present day and early universe.
 
\end{abstract}

\maketitle

 \section{Introduction \label{s-intro}}
  
  Two competing ways  of the cosmic conundrums resolution:\\
{$\bullet$ A. Traditional astrophysical point of view: 
BH formation proceeds by
formation of BHs by stellar collapse or matter accretion
to galactic centres. Unjustified assumptions are demanded, invented just for the case, 
contradicting established physics:
super Eddington accretion, direct collapse, sleeping BHs, but still they are unable to explain many pieces of data.}  
According to my colleague point of view these all are similar to epicycles in ancient astronomy,  in  
the Ptolemaic model. 

{$\bullet$ B. Explanation of observations by primordial BHs, created in the early universe prior to star birth. 
Very well confirmed or supported by "experiment" (observational data),} 
{\bf in particular, predicted log-normal mass spectrum of PBHs, abundant antimatter 
in our Galaxy (in particular, of antinuclej, antistars and even possible star-antistar annihilation)
and existence of supermassive black holes (SMPH) in (almost) empty space or in very small galaxies
in contempory and early universe}\\[1mm]
{ Fritz Zwicky discovered dark matter in the 30th} and was badly criticized by the community.
He referred contemptuously to { “the useless trash in the bulging astronomical journals”}, saying 
{“Astronomers are spherical bastards. 
No matter how you look at them they are just bastards.”}
Still the situation is not as bad as it was in Zwicke's case. Observations  of SMBHs in tiny galaxies and in almost empty space 
leave no room for the direct collapse scenario leaving the only option of primordial black holes.

 During last decade observations made by Hubble Space Telescope (HST),
see e.g. ~\cite{hst-1,hst-2,hst-3}
and very recently by James Webb Space Telescope (JWST)~\cite{jwst-1,jwst-2,jwst-3,jwst-4}
 have led to the surprising conclusion  that the early universe, younger 
than one billion years, is densely populated by well developed galaxies, quasars (supermassive
 black holes), gamma-bursters, and heavy elements (heavier than helium).
 These striking results were taken by the community as absolutely incompatible with the canonical 
$\Lambda$CDM cosmology, especially after release of the JWST data. In fact already observations
of HST could be a sufficient cause for anxiety, not only with respect to the early universe but also to the 
contemporary very old universe almost 15 billion years old. The troubling situation in the
present day universe as well  as in the universe with redshifts $z= 6-10$ are 
summarised in review~\cite{AD-UFN}. 
The state of art is emphatically characterised as crisis in cosmology that 
is believed to hit strong blow to the conventional $\Lambda$CDM picture.

During last decade observations made by Hubble Space Telescope (HST),
see e.g. ~\cite{hst-1,hst-2,hst-3}
and very recently by James Webb Space Telescope (JWST)~\cite{jwst-1,jwst-2,jwst-3,jwst-4}
 have led to the surprising conclusion  that the early universe, younger 
than one billion years, is densely populated by well developed galaxies, quasars (supermassive
 black holes), gamma-bursters, and heavy elements (heavier than helium).
 These striking results were taken by the community as absolutely incompatible with the canonical 
$\Lambda$CDM cosmology, especially after release of the JWST data. In fact already observations
of HST could be a sufficient cause for anxiety, not only with respect to the early universe but also to the 
contemporary very old universe almost 15 billion years old. The troubling situation in the
present day universe as well  as in the universe with redshifts $z= 6-10$ are 
summarized in review~\cite{AD-UFN}. 
The state of art is emphatically characterized as crisis in cosmology that 
is believed to hit strong blow to the conventional $\Lambda$CDM picture.

However, the resolution of the above mentioned problems was suggested 
in our papers~\cite{DS} (DS) and \cite{DKK}  (DKK). 
long before these problems arose. In these works a new
mechanism of massive primordial black hole (PBH) formation was worked out
that could lead to their efficient creation with the masses in the range from a fraction of the solar mass 
up to billion solar masses. 

An essential input of  DS and DKK papers is the suggestion of 
an inverted formation mechanism of galaxies and their central black holes.
Usually it is assumed that supermassive BHs (SMBHs), that are
observed in centres of all large galaxies, are created by matter accretion
to the density excess in the galactic centre, 
but the estimated necessary time is much longer than the universe age, 
even for the contemporary universe, with the age about 15 billion years,
to say nothing about the 20 times younger universe at $z\sim 10$. 

On the opposite, as it was  conjectured in refs.~\cite{DS,DKK}, supermassive black holes were created first in the 
early universe at prestellar epoch, that's why they are caller primordial, and later they have SEEDED galaxy formation.  

Model DS/DKK is verified by a very good agreement of the calculated log-normal
mass spectrum of PBH with observations and by 
discovery of abundant antimatter population in the Galaxy envisaged according to  DS and DKK.
The model also predicts an early formation of galaxies, seeded by PBH,
quasars (alias SMBH), rich chemistry (heavy elements), and dust in the early universe.

\section{A few words about HST and JWST observations \label{s-HST-JWST}}


The orbit of HST  is at the distance of 570 km from the Earth. The orbit  of JWST is much larger, it is about 
$1.5\times 10^6 $ km. The mirror of HST has diameter equal to 2.4 m, while JWST has 2.7 time larger one
and correspondingly the area of JWST mirror is approximately 7.4 times larger. 
In fig. 1 the images of HST and JWST are presented.

\begin{figure}[htbp]
\begin{center}
\includegraphics[scale=0.40,angle=-90]{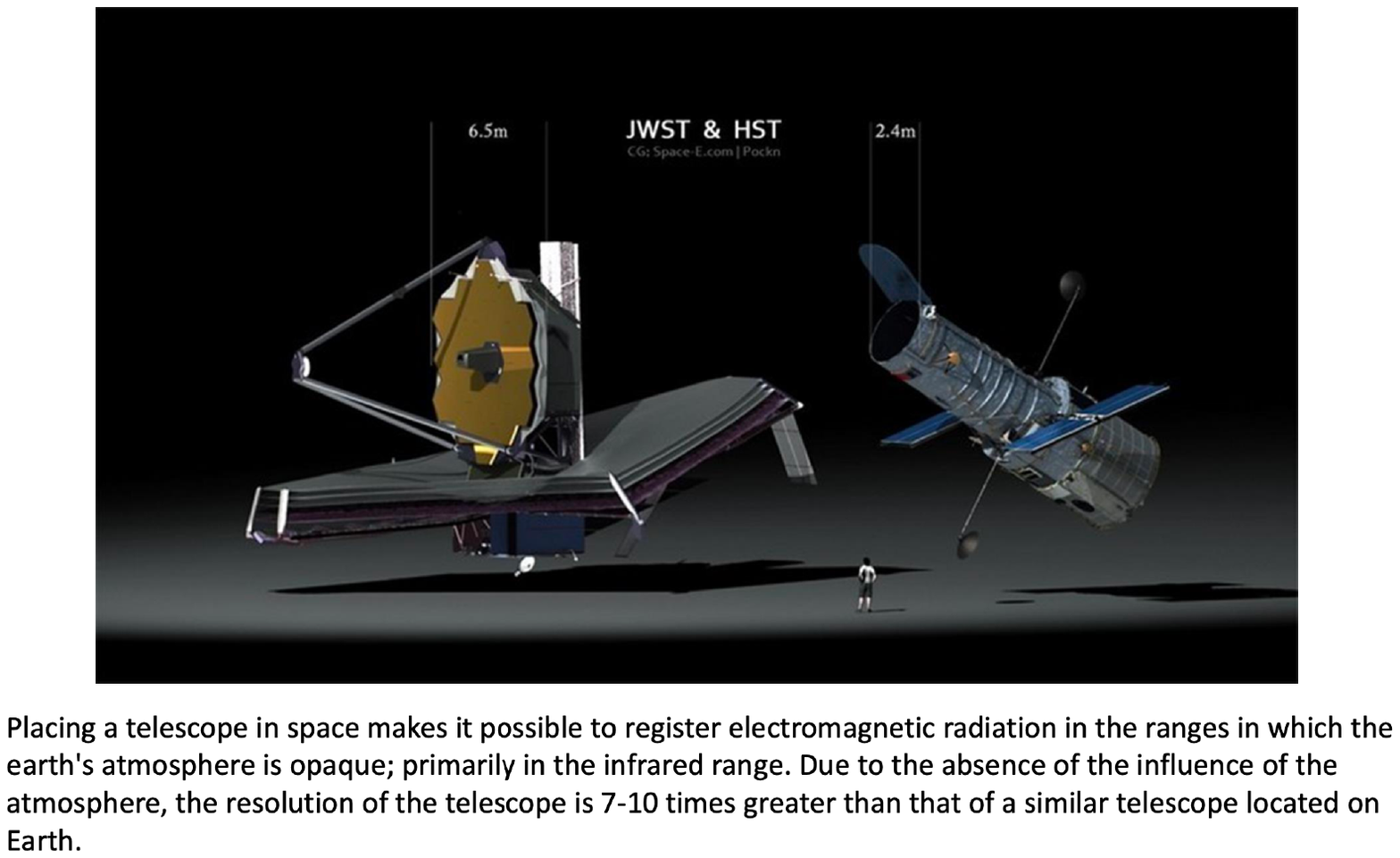}
\caption{Pictures of HST and JWST }
\end{center}
\end{figure}

HST operates in optical wave length range, for example 450 nm, corresponding to blue light. It has also a
possibility to catch the signal in the infrared range with the wave length  0.8-2.5 microns. JWST has high sensitivity
to infrared radiation with the wave length 0.6 - 28,5 micron. It allows to penetrate deep into the early universe,
up to redshifts $z \sim 15$.

Accidentally HST and JWST observed the same galaxy at $z = 12$, see fig. 2. This coincidence is a strong 
argument in favor of the reliable operation of these two very different instruments.
\begin{figure}[htbp]
\begin{center}
\includegraphics[scale=0.34,angle=-90]{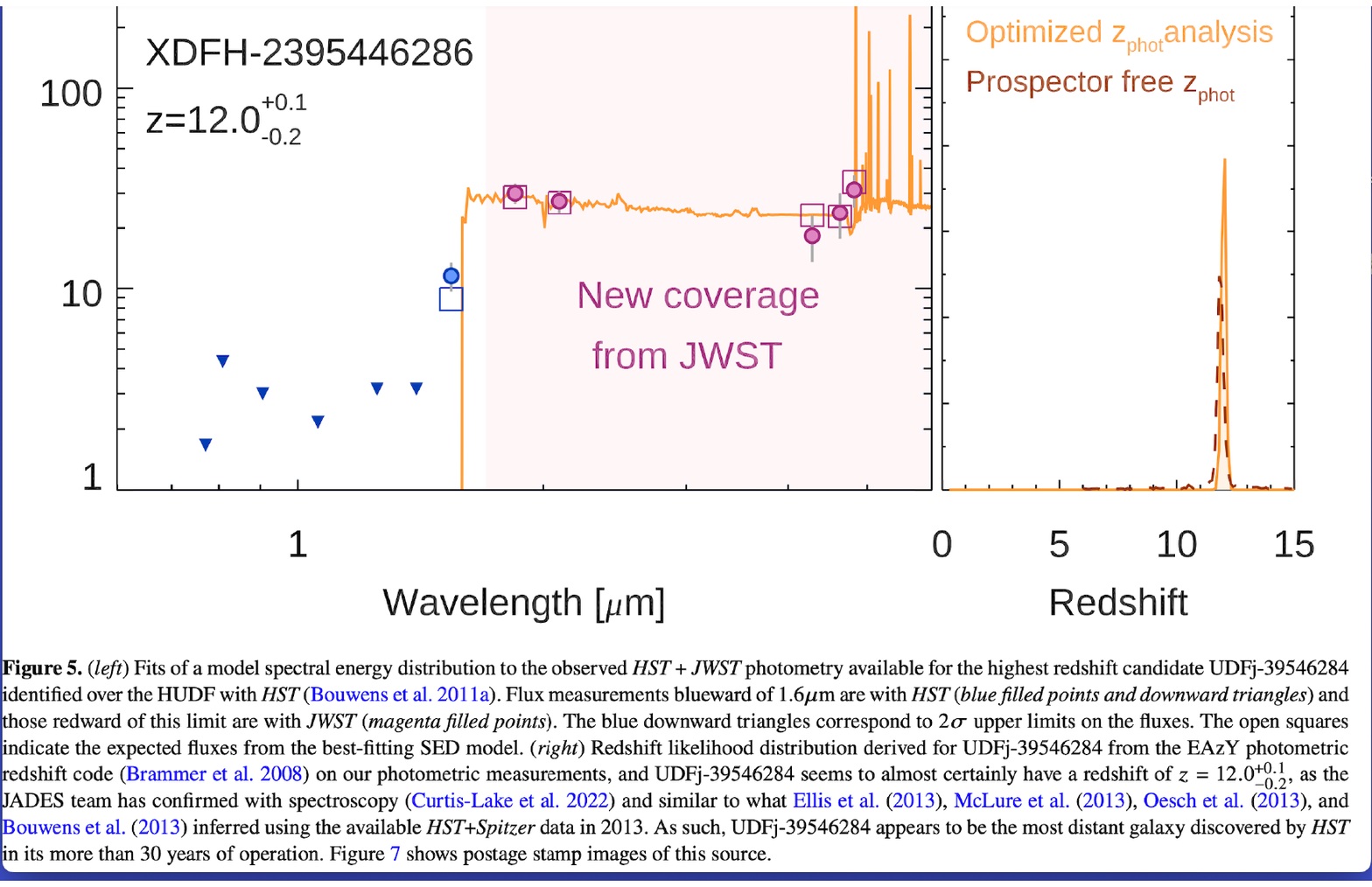}
\caption{JWST and HST common galaxy}
\end{center}
\end{figure}

Comparison of the JWST data and theoretical expectation of the $\Lambda$CDM cosmology is depicted in fig. 3. Theoretical expectations
(colorored dots at $z=15$) are two orders of magnitude below  observations.
\begin{figure}[htbp]
\begin{center}
\includegraphics[scale=0.5,angle=-90]{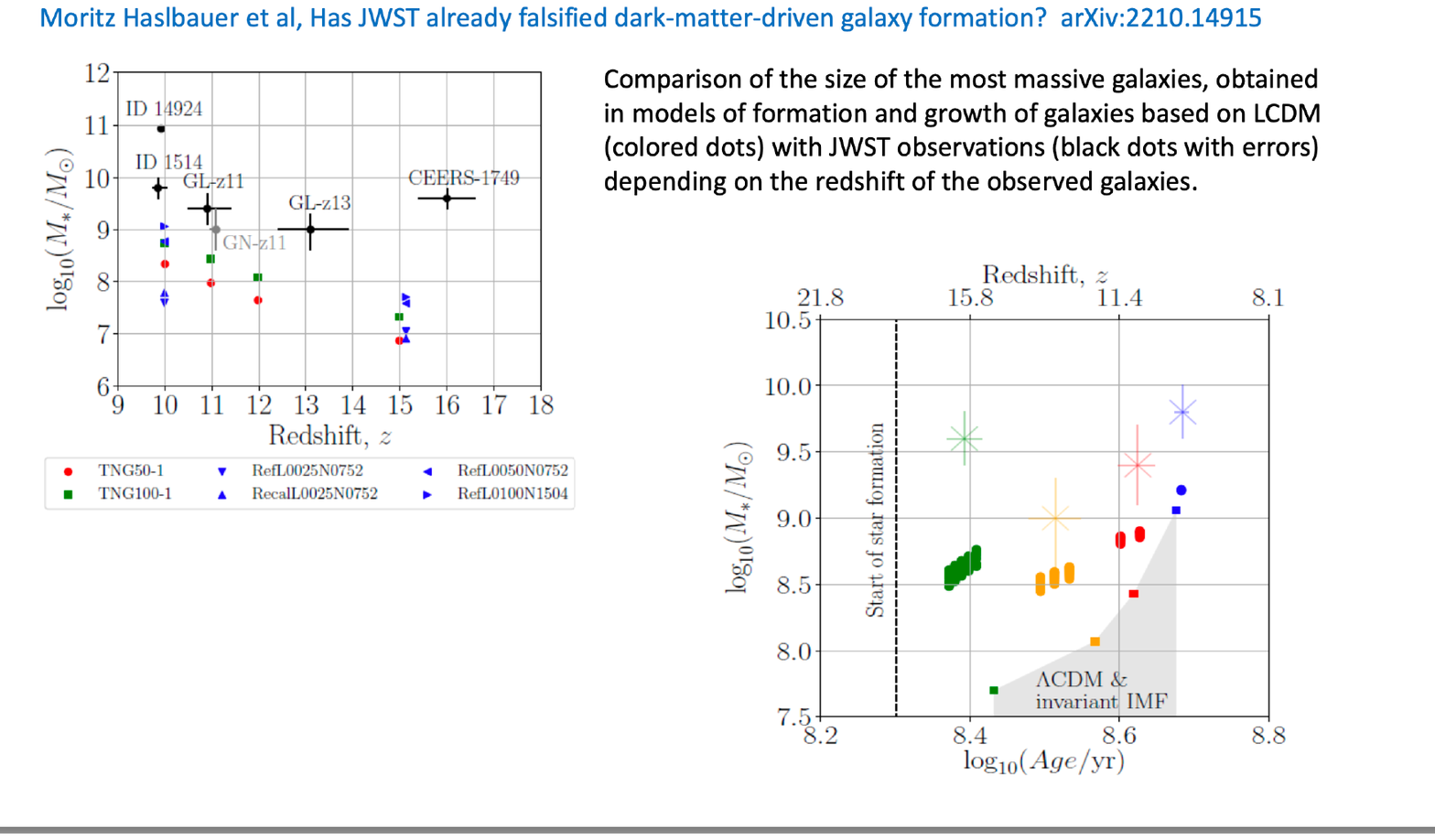}
\caption{Comparison of  JMST observations and and conventional cosmology predictions }
\end{center}
\end{figure}

\section{Spectral measurements and puzzles of the early universe \label{s-spectrum}}

{Only continuum in micron range was measured by JWST
till February. That raised justified doubts on accuracy of 
the redshifts determination of the observed galaxies.}
Now numerous observations of spectra of different elements excellently confirm the early data.
For example according to ref.~\cite{spectrum-z11} 
the JWST NIRCam 9-band near-infrared imaging of the luminous $z=10.6$
galaxy GN-z11 from the JWST Advanced Deep Extragalactic Survey (JADES) proved that 
the  spectral energy distribution (SED) is entirely consistent
with the expected form of the high-redshift galaxy.

In a simultaneous work~\cite{spectrum-z11-2}
the spectroscopy of GN-z11, the most luminous
candidate $z>10$ Lyman break galaxy is presented. The nitrogen lines are clearly observed. 
Quoting the authors: "The spectroscopy confirms that GN-z11 is a remarkable galaxy with extreme
properties seen 430 Myr after the Big Bang."
  
Another example of spectral measurements by a different instrument:
age of most distant galaxy is confirmed with Oxygen observation.
The radio telescope array ALMA (Atacama Large Millimeter Array)
has pin-pointed the exact cosmic age of a distant JWST-identified 
galaxy, GHZ2/GLASS-z12, at 367 million years after the Big Bang~\cite{alma-z12}.
The observations of the a spectral emission line emitted by ionized Oxygen near the galaxy,
red-shifted according to its age in the early universe, confirms the JWST data.  
This data show that the JWST is able to look out to record distances and, quoting the authors,
{\bf heralds a leap in our ability to understand the formation of the earliest galaxies in the Universe.}

A population of red candidate
massive galaxies (stellar mass $ > 10^{10}$ solar masses) at $ 7.4 \lesssim z \lesssim 9.1$, 500–700 Myr after the Big Bang, 
including one galaxy with a possible stellar mass of $\sim 10^{11} M_\odot$, too massive
to be created in so early universe, is observed in \cite{imp-gal}.  Authors conclude that
according to the 'science' it is impossible to create so well developed galaxies. 
 ''May be they are {supermassive {\bf black holes of the kind never seen before.}
That might mean a revision of usual  understanding of black holes.''} 
\vspace{0cm}
\begin{figure}[htbp]
\begin{center}
\includegraphics[scale=0.35,angle=-90]{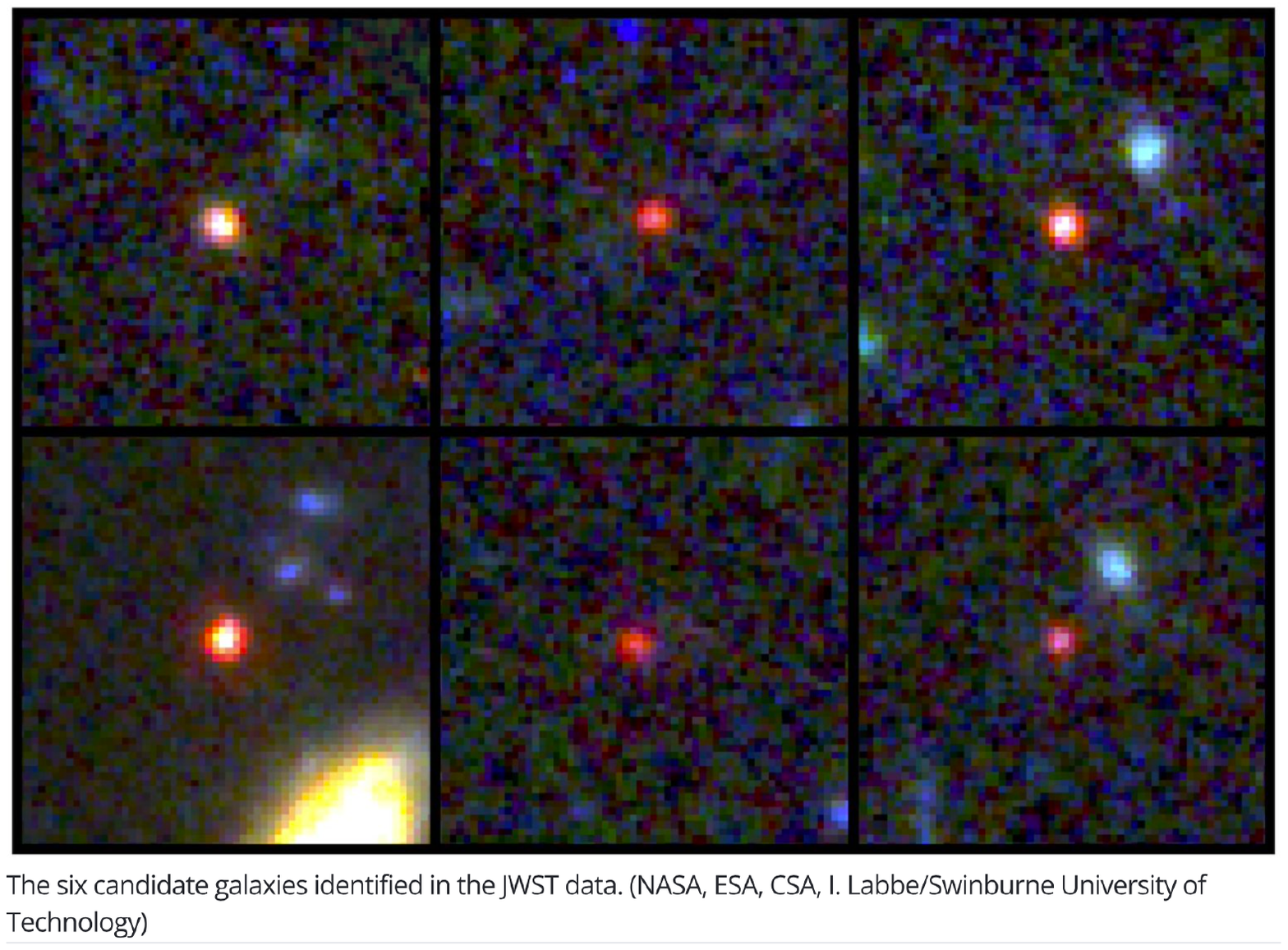}
\caption{Black holes never seen before{?}}
\end{center}
\end{figure}

Clearly these "black holes of the kind never seen before" nicely fit the assertion that they are primordial, as suggested 
in refs.~\cite{DS,DKK}.

Recent observation by ALMA~\cite{alma-6.8} of an extremely massive reionization-era galaxy with $M_* = 1.7 \times 10^{11} M_\odot $
at z = 6.853 with active galactic nuclei (AGN) that have huge luminosity suggests that this object is powered by 
${ \sim 1.6 \times  10^9 M_\odot }$ black hole if accreting closely to the Eddington limit. It is nearly impossible to create so 
massive BH in  the early universe.  But supermassive primordial black hole could easily feed such monster

\section{Rich chemistry in the early universe \label{sec-chem}}

According to the standard lore, light elements, deuterium, helium, and tiny traces of lithium are created from the primordial protons and neutrons 
in the early universe roughly during the first 100 seconds. This process is called big bang nucleosynthesis (BBN). 
Heavier elements, the so called metals (everything heavier than helium-4 are called
metals in astrophysics) are created through stellar nucleosynthesis. Next, supernova explosions populate interstellar medium with
metals. 

Unexpectedly  high abundances of heavy elements (high metallicity) are observed in the very early universe by HST and JWST.
For example, as reported in ref.~\cite{peng-metals},  a study of a strongly lensed galaxy SPT0418-47 revealed its
mature metallicity, amounts of elements heavier than helium and hydrogen, such as carbon, oxygen and nitrogen.
According to the estimate of the team the amount is 
comparable to that of the sun, which is more than 4 billion years old and inherited 
most of its metals from previous generations of stars that had 8 billion years to build them up.
Analysis using optical strong line diagnostics suggests that galaxy SPT0418-SE has
 near-solar elemental abundance, while the ring appears to have supersolar metallicity O/H and N/O. 

One more example of well developed chemistry, that demands too long evolution if produced by 
the conventional mechanism is presented in ref.~\cite{cameron}.
Observations of GN-z11 with JWST/NIRSpec disclosed numerous oxygen,
carbon, nitrogen, and helium emission lines at $z=10.6$. 
The data prefers (N/O), greater than 4 times solar and
the derived $C/O \approx 30$ solar.
{Nitrogen enhancement in GN-z11 cannot be explained by enrichment from metal-free
Population III stars.}
The suggested explanation is that yields from runaway stellar collisions in
a dense stellar cluster or a tidal disruption event provide promising solutions
to give rise to these unusual emission lines at $z=10.6$, and explain the
resemblance between GN-z11 and a nitrogen-loud quasar. 

High abundances of heavy elements may be a result of BBN with baryon-to-gamma ratio close to unity,
as it takes place in the  DS~\cite{DS} and DKK~\cite{DKK} model,
see below. 

The hypothesis pioneered by DS~\cite{DS}  and DKK~\cite{DKK}, 
that  galaxy formation is seeded by SMBH allows to understand
the presence of SMBH in all large and several small galaxies accessible to observation.
{This mechanism explains how  the galaxies observed by
JWST in the very young universe might be created.} 
Presently it is rediscovered in several recent works.

As is stated in ref.~\cite{liu},  
the recent observations with JWST have identified several bright galaxy
candidates at $z\gtrsim 10$, some of which appear unusually massive (up to
$\sim 10^{11}\ \rm M_{\odot}$). Such early formation of massive galaxies is
difficult to reconcile with standard $\Lambda\rm CDM$ predictions
demanding very high star formation efficiency (SFE), possibly even in excess of the
cosmic baryon mass budget in collapsed structures. With an idealized analysis
based on linear perturbation theory and the Press-Schechter formalism, 
the observed massive galaxy candidates can be explained,
with lower SFE than required in $\Lambda\rm CDM$, 
 if structure formation is accelerated by
massive ($\gtrsim {10^{9}\ \rm M_{\odot}}$) PBHs 
that enhance primordial density fluctuations.

{Observations made by JWDST (and HST) of high-redshift quasars reveal that many supermassive black holes were in place 
less than 700 Million years after the Big Bang. }
In particular, in ref.~\cite{bogdan-1} the 
detection of an X-ray-luminous quasar powered by SMBH with the mass 
 $\sim 4 \times 10^7 M_\odot$ in a gravitationally-lensed galaxy,
 identified by JWST at $z \approx 10.3$, is reported. As is stated by the authors,
 this mass is comparable to the inferred stellar mass of its host galaxy, in contrast to 
the  usual examples from the local universe where mostly
the BH mass is $\sim 0.1$\% of the host galaxy's stellar mass. 
The combination of such a high BH mass and large BH-to-galaxy stellar mass 
ratio $\sim 500$ Myrs after the Big Bang 
is consistent with a picture wherein 
{ such BHs originated from heavy seeds.}
{Let stress again, that this detection 
suggests that early supermassive black holes originate from} {\bf heavy seeds.} 

{ However, the origin of the first BHs, that started the seeding, remains a mystery.}
According to the authors. the seeds of the first BHs are postulated to be either light i.e., $(10-100) M_\odot$
remnants of the first stars
{or heavy i.e., $(10^4 - 10^5) M_\odot$,} originating from direct collapse of gas clouds.
The latter hypothesis is questionable, but a supermassive primordial black hole would perfectly work.

In a subsequent paper~\cite{seed-2} a 
support to the heavy seeding channel for the formation of supermassive BHs within the first billion years of cosmic evolution is
also proposed. As is mentioned in this work,
"the James Webb Space Telescope is now detecting early black holes (BHs) as they originate from 
{seeds}  to supermassive BHs. Recently Bogdan et al~\cite{bogdan-1} reported the detection of an X-ray luminous 
supermassive black hole, UHZ-1, 
with a photometric redshift at $z > 10$. Such an extreme source at this very high redshift provides new insights on {\bf seeding} 
and growth models for BHs {given the short time available for formation and growth.}
The resulting ratio of $M_{BH}/M^*$ remains two to three orders of magnitude higher than local values, 
thus lending support to the heavy {seeding} channel for the formation of supermassive BHs 
within the first billion years of cosmic evolution.

\section{Seeding of galaxy formation by PBH \label{s-seeding}}

\subsection{Seeding of galaxies \label{ss-seed-gal|}}

The hypothesis pioneered by DS~\cite{DS}  and DKK~\cite{DKK}, 
that  galaxy formation is seeded by SMBH allows to understand
the presence of SMBH in all large and several small galaxies accessible to observation.
{This mechanism explains how  the galaxies observed by
JWST in the very young universe might be created.} 
Presently it is rediscovered in several recent works.

As is stated in ref.~\cite{liu},  
the recent observations with JWST have identified several bright galaxy
candidates at $z\gtrsim 10$, some of which appear unusually massive (up to
$\sim 10^{11}\ \rm M_{\odot}$). Such early formation of massive galaxies is
difficult to reconcile with standard $\Lambda\rm CDM$ predictions
demanding very high star formation efficiency (SFE), possibly even in excess of the
cosmic baryon mass budget in collapsed structures. With an idealized analysis
based on linear perturbation theory and the Press-Schechter formalism, 
the observed massive galaxy candidates can be explained,
with lower SFE than required in $\Lambda\rm CDM$, 
 if structure formation is accelerated by
massive ($\gtrsim {10^{9}\ \rm M_{\odot}}$) PBHs 
that enhance primordial density fluctuations.

{Observations made by JWDST (and HST) of high-redshift quasars reveal that many supermassive black holes were in place 
less than 700 Million years after the Big Bang. }
In particular, in ref.~\cite{bogdan-1} the 
detection of an X-ray-luminous quasar powered by SMBH with the mass 
 $\sim 4 \times 10^7 M_\odot$ in a gravitationally-lensed galaxy,
 identified by JWST at $z \approx 10.3$, is reported. As is stated by the authors,
 this mass is comparable to the inferred stellar mass of its host galaxy, in contrast to 
the  usual examples from the local universe where mostly
the BH mass is $\sim 0.1$\% of the host galaxy's stellar mass. 
The combination of such a high BH mass and large BH-to-galaxy stellar mass 
ratio $\sim 500$ Myrs after the Big Bang 
is consistent with a picture wherein 
{ such BHs originated from heavy seeds.}
{Let stress again, that this detection 
suggests that early supermassive black holes originate from} {\bf heavy seeds.} 

{ However, the origin of the first BHs, that started the seeding, remains a mystery.}
According to the authors. the seeds of the first BHs are postulated to be either light i.e., $(10-100) M_\odot$
remnants of the first stars
{or heavy i.e., $(10^4 - 10^5) M_\odot$,} originating from direct collapse of gas clouds.
The latter hypothesis is questionable, but a supermassive primordial black hole would perfectly work.

In a subsequent paper~\cite{seed-2} a 
support to the heavy seeding channel for the formation of supermassive BHs within the first billion years of cosmic evolution is
also proposed. As is mentioned in this work,
"the James Webb Space Telescope is now detecting early black holes (BHs) as they originate from 
{seeds}  to supermassive BHs. Recently Bogdan et al~\cite{bogdan-1} reported the detection of an X-ray luminous 
supermassive black hole, UHZ-1, 
with a photometric redshift at $z > 10$. Such an extreme source at this very high redshift provides new insights on {\bf seeding} 
and growth models for BHs {given the short time available for formation and growth.}
The resulting ratio of $M_{BH}/M^*$ remains two to three orders of magnitude higher than local values, 
thus lending support to the heavy {seeding} channel for the formation of supermassive BHs 
within the first billion years of cosmic evolution.

\begin{figure}[htbp]
\begin{center}
\includegraphics[scale=0.7,angle=-90]{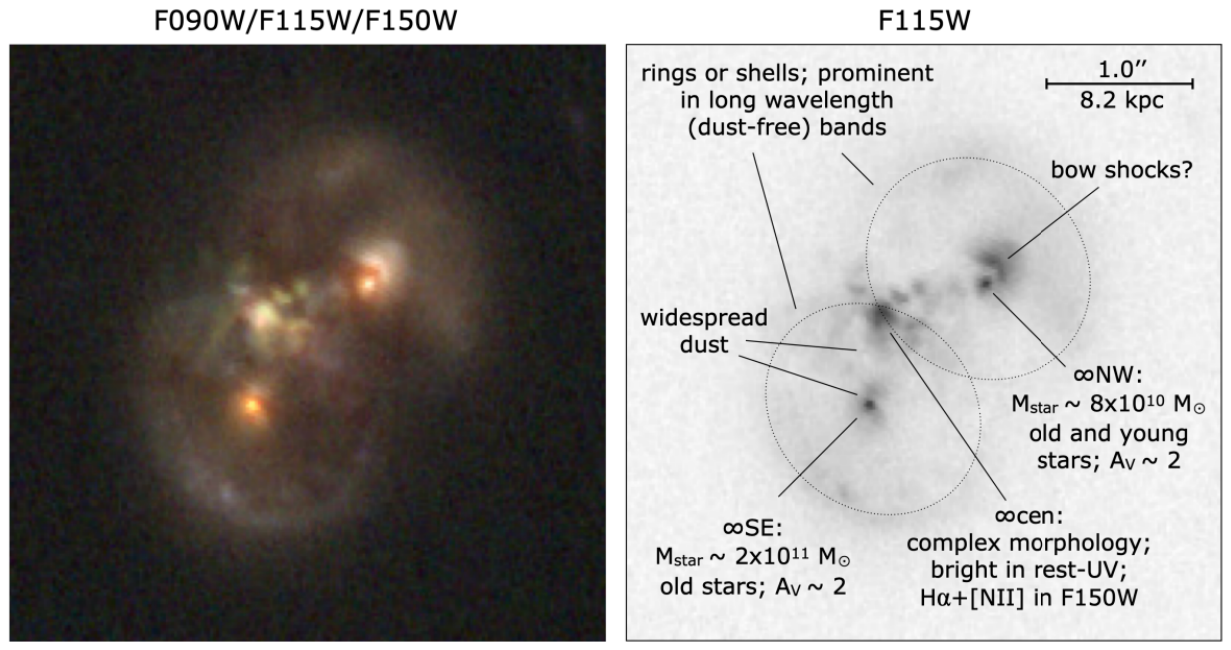}
\caption{Triple black hole}
\label{tripBH}
\end{center}
\end{figure}

A very interesting result  is reported in ref.~\cite{tri-BH}. Two galaxies with central supermassive black holes are observed 
and a third SMBH between them, see fig.~\ref{tripBH}. According to the authors the middle SMBH was created by direct collapse
of cold interstellar gas. It looks quite unnatural. The assumption that the BH in the middle is primordial one does not
suffer from any inconsistencies.

\subsection{ Seeding of globular clusters and dwarf galaxies \label{seed-dwarf}}

The idea of seeding of globular clusters and dwarf galaxies by primordial black holes was worked out in ref.~\cite{AD-KP-dwarfs}.
Primordial IMBHs with masses of a few thousand solar mass  can explain  their formation, poorly understood otherwise.
In the last several years such IMBHs inside globular clusters are observed. Similar features are true for dwarfs.
In particular the seeding of dwarfs by intermediate mass BHs
is confirmed by the recent data. For instance in the dwarf galaxy
SDSS J1521+1404 the BH is discovered with the mass 
$M \sim 10^5 M_\odot$~\cite{seed-dwarf-1}.
For the first time, astronomers have spotted evidence of a 
{pair of dwarf galaxies featuring GIANT black holes on a collision course with each other. 
In fact, they haven't just found just one pair – they've found two.}

Another recent example~\cite{dwarf-2} of intermediate-mass black holes is the 
finding of episodic, large-scale and  powerful jet activity in a dwarf galaxy
  SDSS J090613.77+561015.2.
This can be explained by an intermediate-mass black hole (IMBH) with a mass of 
$M_{BH} = 3.6^{+5.9}_{-2.3} \times 10^5 M_{\odot}$, 
Such huge black hole surely could not be created 
by accretion but vice versa might seed the  formation of the dwarf.

\section{Peculiar stars in the Galaxy \label{s-pec-stars} }

\subsection{Ancient stars \label{mature}}

A discovery of primordial stars in globular cluster M92~\cite{prim-stars} was recently announced. The 
absolute age of the globular cluster M92 was evaluated and found to be practically equal to the universe
 age, ${t_{M92} = 13.8 \pm 0.75 }$ Gyr. As it is  stated in the paper, possibly these  stars
 came to us from JWST epoch or even from the earlier one.

Similar declaration of  pristine 
stars in the Galaxy was made almost at the same time~\cite{pristine}.
An international team of researchers, Pristine Inner Galaxy Survey (PIGS) team,
has obtained the largest set of detailed observations yet of the 
oldest stars in the center of our Galaxy, the Milky Way. 
Some of the stars that were born in the first billion years after the Big Bang 
are still around today.

In fact, extremely old stars in the Galaxy were discovered considerably earlier. As it is asserted in ref.~\cite{old-star-1},
new more accurate methods of determination of stellar ages led to discovery of surprisingly old stars.
Employing thorium and uranium abundances in comparison with each other and with
several stable elements the age of metal-poor, halo star BD+17o 3248 was estimated as
$13.8 \pm 4$ Gyr. For comparison the age of inner halo of the Galaxy is 11.4 ± 0.7 Gyr~\cite{halo-age} 

The age of a star in the galactic halo, HE 1523-0901, was estimated to be about 13.2
Gyr. First time many different chronometers, such as the U/Th, U/Ir, Th/Eu and
Th/Os ratios to measure the star age have been employed~\cite{old-star-2}.

And now, the most surprising star which is older than the Universe. 
Metal deficient high velocity subgiant in the solar neighborhood HD 140283 has the age
14.46 ± 0.31 Gyr [28]. The determined central value of the age exceeds the universe age
by two standard deviations if the Hubble parameter is low, H = 67.3 (according to the
CMB analysis) and $t_U = 13.8$ Gyr, if H = 74 (according to the traditional methods),
and $t_U = 12.5$ Gyr. The age of this star exceeds the universe age more than by 10 $\sigma$.

In our model~\cite{DS,DKK} not only primordial black holes could be formed but, if  the
bubbles with high baryon-to-photon ratio are not sufficiently massive, compact stellar kind objects 
could be created. Such "stars" might look older than they are because they would be enriched with
heavy elements mimicking larger age.

\subsection{Fast moving stars \label{ss-fast}}

Several stars are discovered in the Galaxy with unusually high velocity  much larger than the galactic virial
velocity, that is about 200 km/sec. There are several very fast pulsars in the Galaxy, but their origin is evident. Pulsars
are the results of supenova explosions and a small angular asymmetry in the emitted
radiation could create a strong kick, which would accelerate a pulsar up to $10^3$ km/sec.
The observed fast stars look normal, except of very high velocity, about 500 km/sec. 

In ref.~\cite{vennes} a a discovery of a low mass white dwarf, LP 40-365, was reported,
that travels at a velocity greater than the Galactic
escape velocity and whose peculiar atmosphere is dominated by intermediate mass elements. 
According to the authors
these properties suggest that it could be the predicted leftover remains from a type Iax supernova.
On the other hand, it can
naturally be a primordial star with high initial abundances of heavy elements.

Let us mention several more discoveries of other high velocity stars in the
Galaxy~\cite{hattori,marchetti}. The authors argue that these stars could be accelerated by a population of
intermediate mass black holes (IMBHs) in 
Globular clusters, if there is sufficient number of IMBHs. So many IMBHs
were not expected but the recent data reveal more and more of them in contrast to
conventional expectations and in agreement with refs.~\cite{DS,DKK}.

As it is noted in ref.~\cite{ruffini}  observations~
of stellar remnants linked to Type Ia and Type Iax supernovae are necessary to fully understand their progenitors and explain the origin of their high
speed. Multiple progenitor scenarios predict a population of kicked donor remnants and partially-burnt primary remnants, both moving with 
relatively high velocity. But only a handful of examples consistent with these two predicted populations have been observed. 
It is reported in ref.~\cite{ruffini} that the likely first known example of an unbound white dwarf 
that is consistent with being the fully-cooled 
primary remnant  to a Type Iax supernova is LP 93-21. The candidate, LP 93-21, is travelling with a 
galactocentric velocity of $v_{gal}  \approx 605$ km/sec, and is gravitationally unbound to the Milky Way. 
The authors claim ruling out its extragalactic origin. 
The Type Iax supernova ejection scenario is consistent with its peculiar unbound trajectory, given the observed anomalous elemental abundances. 
This discovery reflects recent models that suggest stellar ejections likely occur often. 

Let us repeat here that extragalactic primordial star presumably populating the galactic halo, according to assertion of papers~\cite{DS,DKK},
very well fits the observations made in ref.~\cite{ruffini}

\subsection{Stars with unusual chemistry \label{ss-chem}}

An unusually red star was observed in planetary system through microlensing
event MOA-2011-BLG-291~\cite{bennet}.
The host star and planet masses are estimated as $ M_{host} = 0.15^{+027}_{-0.10} M_\odot$
 and $m_{planet}  = 18^{+34}_{-12} M_\oplus$.
The source star that is redder (or brighter) than the bulge main sequence.  
The favoured interpretation by the authors  is that the source star is a lower main sequence star at a distance of 
$4.9 \pm 1.3$ kpc in the Galactic disk.
According to the authors, the life-time of main sequence star with
the solar chemical content is larger than the universe age already for $M < 0.8 M_\odot$. 
It implies the primordial origin of the registered star with already evolved chemistry. 

May it be a primordial helium star? There could be stars dominated by helium, even purely helium stars, in our scenario~\cite{DS,DKK}.

\section{Pulsar humming \label{s-humm}}

If a pulsar moves in any way, orbiting around a star, the relative motion of the pulsar causes the pulses to shift slightly. 
These shifts can be measured with extreme accuracy. 
{The observations are so precise, pulsars were used to measure 
the orbital decay of binary systems as indirect evidence of gravitational waves long before they are observed directly.}

{Unexpectedly high number of SMBH binaries} are presumably
observed through distortion of the pulsar timing by
emission of gravitational waves~\cite{puls-hum}.
The NANOGrav 15 yr data set shows evidence for the presence of a low-frequency gravitational-wave background. 
While many physical processes can source such low-frequency gravitational waves, but most natural possibility seems to be that
 the signal as coming from a population of supermassive black hole (SMBH) binaries distributed throughout the Universe~\cite{puls-hum}.

It is difficult to explain such huge number of SMBH binaries.   
However, this can be naturally expected if these SMBHs are primordial.

\section{Possible types of black holes \label{s-BH-types}}

\subsection{BH classification by mass \label{ss-BHbyMass}}

There is the following conventional division of black holes by their masses:\\
1. Supermassive black holes (SMBH): $ M = (10^6 -10^{10}) M_\odot$ (the record mass is about $10^{11} M_\odot$).\\
{2. Intermediate mass black holes (IMBH):  $ M = (10^2 -10^{5}) M_\odot$.}\\
{ 3. Solar mass black holes: masses from a fraction of $M_\odot$ up to $100 M_\odot$}.\\
{The origin of most of these BHs is unclear in the traditional approach, 
except maybe of the BHs with masses of a few
solar masses, that might be astrophysical}.
{Highly unexpected was a great abundance of IMBH which are copiously appearing in observations
during last few years.}

{The assumption that (almost) all black holes in the universe are primordial strongly reduce or 
even eliminate the tension between the data and expected numbers of black holes.}

\subsection{BH classification by formation mechanism \label{ss-BHbyFormation}}

{\bf 1. Astrophysical  black holes,}
{created by the collapse of a star that exhausted its nuclear fuel.}
The expected masses should start immediately above the neutron star mass, i.e. about
{ ${3M_\odot}$, but noticeably below $100 M_\odot$.}
Instead we observe that the BH mass spectrum in the galaxy has maximum at
 { ${M \approx 8 M_\odot}$ with the width  ${ \sim(1-2) M_\odot }$.  } 
 {The result is somewhat unexpected but an explanations in the conventional astrophysical frameworks might be  
 possible. } 
 { Recently LIGO/Virgo discovered black holes with masses close to $100 M_\odot$. }
 Their astrophysical origin was considered unfeasible due to huge mass loss in the process of collapse. 
 Now some, quite exotic, formation mechanisms are suggested. \\
{\bf 2. BH  formed by accretion on the mass excess in the galactic center.}
{In any large galaxy there exists a supermassive BH (SMBH) at the center, with masses
varying from several millions of $M_\odot$ (e,g, Milky Way) up to almost hundred billions $M_\odot$.}
{However, the conventional accretion mechanisms are not efficient enough to create such
monsters during the universe life-time, ${ t_U \approx 14.6 }$ Gyr. At least 10-fold longer time
is necessary,} {to say nothing about SMBH in 10 times younger universe.}\\
{\bf  3. Primordial black holes (PBHs) created during pre-stellar epoch.}
{The idea of the primordial black hole (PBH)  i.e. of black hole which have been 
formed  in the early universe prior to star formation,
was first put forward by Ya.B. Zeldovich and I.D. Novikov~\cite{ZN-PBH}.}
{According to their arguments, if the 
density contrast in the early universe inside the bubble with radius equal to the cosmological horizon 
might accidentally happen to be large, {${\delta \rho /\rho \approx 1}$,} then
that piece of volume would be inside its gravitational radius i.e. it became  a PBH, that
decoupled  from the cosmological expansion. }
Subsequently this mechanism was elaborated by S. Hawking~\cite{SH-PBH} and B.Carr and S.Hawking~\cite{Carr-SH}

\section{PBH and inflation \label{s-BH-infl} }

In earlier works the predicted masses of PBH were quite low, more or less equal to the mass of the matter inside cosmological horizon
at the moment ob PBH formation. {Inflation allowed for formation of PBH with very large masses.
It was first applied to PBH creation by Dolgov and Silk~\cite{DS}, 
and a year later by Carr, Hilbert, and Lidsey~\cite{carr-infl},
and soon after that by Ivanov, Naselsky, and Novikov~\cite{INN}.

{Presently inflationary mechanism of PBH production  is commonly used. It allows to create 
PBH with very high masses, but the predicted spectrum is multi-parameter one and  quite complicated.}
{The only exception is the log-normal spectrum of  refs.~\cite{DS, DKK}
which is verified by observations in excellent agreement:} 
\be 
\frac{dN}{dM} = \mu^2 \exp[-\gamma \ln^2 (M/M_0)], 
\label{dn-dm}
\ee
where the central mass $M_0$ is theoretically predicted to be equal to the horizon mass at the moment of the QCD phase transition
from free quarks to confinement phase consisting of baryons~\cite{AD-KP}. Assuming that the QCD phase transition takes place at 100 MeV,
we find that $M_0 \approx 10 M_\odot$. Usually the temperature of the phase transition is determined by lattice calculations which give 
a close result. More accurate calculations with an account of nonzero value of the quark chemical potential, $\mu_q$ were performed 
in ref.~\cite{arefyeva}, where it was shown that with the necessary value of $\mu_q$ the horizon mass is $17 M_\odot$ that is 
surprisingly close to the best fit value found from the chirp mass distribution of the coalescing black holes, see fig. \ref{fig-EDF} below.

\section{Black Dark Matter \label{black-DM}}
 
The first suggestion PBH might be dark matter "particles"  was made by
S. Hawking in 1971~\cite{SH-BH}
and later by G. Chapline in 1975~\cite{chaplin}, who simply noticed that low mass
PBHs might be abundant 
in the present-day universe and their energy density could be comparable to the energy density of dark matter. 
In the latter paper the scale independent spectrum of cosmological perturbations was assumed, thus leading to the
flat  PBH mass spectrum in log interval:
\be
{dN = N_0 (dM/M) }
\ee
with maximum mass { ${M_{max} \lesssim 10^{22}  }$ g,} which hits the allowed mass range.
{The next proposal of BH-dominated dark matter  was made in ref.~\cite{DS} that even was
contained in the title
"Baryon isocurvature fluctuations at small scales and  {\bf baryonic dark matter,}"
with much larger and more realistic black hole masses, close to $10M_\odot$.

\subsection{Bounds on BH energy density \label{ss-DM-limirts}}

The constraints on the density of black holes were reviewed by Carr and Kuhnel~\cite{BH-limits} and the results are presented 
in Fig.  5
{for monochromatic mass spectrum of PBHs.} 
\begin{figure}[htbp]
\begin{center}
\includegraphics[scale=0.65]{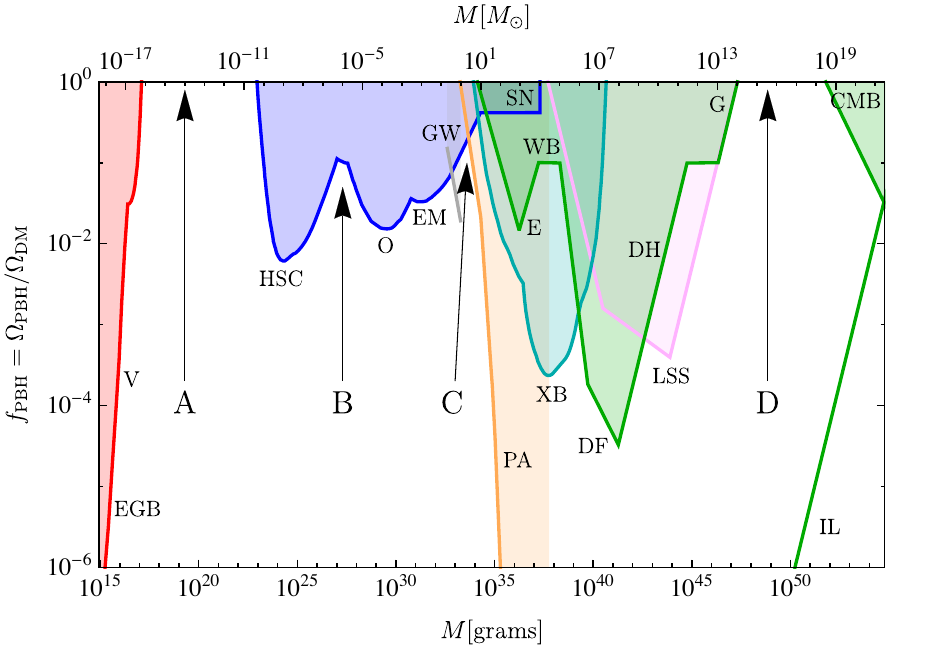}
\caption{
Constraints on $f( M )$ 
			for a { monochromatic} 
			mass function,
			from evaporations (red), 
			lensing (blue), 
			gravitational waves (GW) (gray),
			dynamical effects (green), 
			accretion (light blue), 
			CMB distortions (orange) and 
			large-scale structure (purple).
			Evaporation limits from the extragalactic gamma-ray background (EGB), 
			the Voyager positron flux (V) and 
			annihilation-line radiation from the Galactic centre (GC). 
			Lensing limits from 
			microlensing of supernovae (SN) and of stars in M31 by Subaru (HSC), 
			the Magellanic Clouds by EROS and  MACHO (EM) 
			and the Galactic bulge by OGLE (O). 
			Dynamical limits  from 
			wide binaries (WB), 
			star clusters in Eridanus II (E),
			halo dynamical friction (DF), 
			galaxy tidal distortions (G),
			heating of stars in the Galactic disk (DH) 
			and the CMB dipole (CMB). 
			Large scale structure constraints(LSS). 
			Accretion limits from X-ray binaries (XB) and 
			Planck measurements of CMB distortions (PA).
			The incredulity limits (IL) correspond to one PBH per relevant environment 
			(galaxy, cluster, Universe). 
			{There are four mass windows (A, B, C, D) 
			in which PBHs could have an appreciable density. }
}
\end{center}
\end{figure}


\subsection{Lifting the bounds on the black hole fraction in dark matter \label{ss-lift-limit} }

At the beginning of this section it would be proper to quote Bernard Carr's words at 2019 : "all limits are model dependent and have caveats."

There are several papers, where authors looking and finding ways to eliminate or weaken the limits on the number density of black holes.

In ref.~\cite{boehm} it is argued that primordial black holes in the mass range $(30-100) M_\odot$ could be the dark matter carriers
since they might escape microlensing and cosmic microwave background constraints. They are however subject to the constraints from
the binary merger rate observed by the LIGO and Virgo experiments. The authors argue that in realistic situation the masses of black holes
in expanding universe depend upon time and this leads to a suppression of binary formation. Hence they conclude that this effect
reopens the possibility for dark matter in the form of LIGO-mass PBHs.

In ref.~\cite{frampton} the authors have opened the window for  PBH with masses in the range
$(10^2-10^5)M_\odot$ to make full or significant contribution to cosmological dark matter.
They claim that the derivation of the accepted bound that excluded considerable contribution of PBH in this mass range
is based on oversimplified accretion model.

As is argued in ref.~\cite{rubin}, PBHs can form clusters.
Dynamical interactions in PBH clusters offers additional channel for the orbital energy dissipation thus increasing the merging 
rate of PBH binaries, and the 
constraints on the fraction of PBH in dark matter can be weaker than that  
obtained by assuming a homogeneous PBH space distribution.

 A recent analysis performed in paper~\cite{eroshenko} permits to conclude that a possible clustering of PBH could significantly reduce
 efficiency of the merger process and the final rate of gravitational wave bursts in some parameter range. As a result 
 the fraction of PBH in dark matter could be as large as unity without distortion of  LIGO/Virgo observational data.
 
The last decisive blow (by importance and not by order)  to the restriction on the the upper bounds on BH density
has been made in ref.~\cite{Bellomo} where it is shown that 
{ a window that is closed for a Monochromatic Mass Distribution (MMD) 
can open for an MMD, in particular for log-normal mass spectrum.}

Let us also ention ref.~\cite{KF} where 
it is argued that there a window in the mass range $10^{-10} M_\odot$ to $10^{-8} M_\odot$ 
to accommodate for 100\% PBH dark matter. 



 \section{Observations of black holes \label{s-BHobserv}}

A possibility of black hope existence was ingeniously  discovered in 1783 by John Michell, an English country parson, famous for many
 other discoveries in physics. He noticed that there could 
  be stellar bodies having the second cosmic velocity larger than the speed of light. Since such objects neither shine nor
 reflect light, it would be impossible to observe them directly. Michell called such, not emitting light stars as "dark stars". According to his 
 understanding a single dark star would be invisible, but if a double system of a dark and a usual star is formed, one may
identify dark star observing the other one rotating around "nothing". This is one of possible ways to observe black holes at the present
time.
 
However, all that happened to be not absolutely true, or possibly even entirely wrong.
{BHs evaporate and shine (Hawking radiation), though nobody  yet  saw it.}
{The most powerful sources of radiation (quasars) are supermassive black holes, point-like 
objects radiate as thousands galaxies through ultrarelativistic particle collision in the process of matter accretion.}
{Near-solar mass BHs are observed through X-rays from accreting surrounding matter.}
Black holes may act as gravitational lenses, that's how MACHOs and some other BHs are discovered.
Observation of the stellar motion around supposed black hole permits to identify the latter as e.g. supermassive black hole in our Galaxy
was discovered.
{All these methods only allow to determine the mass inside central volume. According to theory of
General Relativity, a huge mass in a small volume must form a black hole.} 
 However, strictly speaking BH existence is not proven by all these methods.
 
 The first direct proof of black hole existence was the 
registration of gravitational waves from a pair of coalescing massive bodies by LIGO/Virgo/Kagra. 
The data explicitly shows that the the coalescence is indeed between two 
black holes, because the best fit to the form of the signal is achieved 
under assumption of the Schwarzschild metric
that according to GR describes non-charged and (almost) non-rotating black hole.
The observations permit to determine the masses of two coalescing BHs, their spins, and the mass of the final black hole.

\section { Gravitational waves from BH binaries}

\subsection{Are the GW sources primordial BHs? \label{ss-gw-pbh}}

As it is argued e.g. in paper~\cite{BDPP}, discovery of
gravitational waves (GW) by LIGO interferometer strongly indicates that the sources of GW are primordial black holes.
 In fact here is general agreement between several groups of theorists, that the gravitational waves discovered by
LIGO/Virgo interferometers originated from PBH binaries. We discuss this issue  here 
following our paper~\cite{BDPP}. There are three features that indicate that the sources of
GWs should most naturally be primordial black holes:\\
{\bf 1. Origin of heavy BHs  (with masses ${\sim 30 M_\odot}$).} To form so heavy BHs, the progenitors should have 
${M > 100 M_\odot}$ and  a low metal abundance to avoid too much
mass loss during the evolution. Such heavy stars might be present in
young star-forming galaxies but they are not observed in the necessary amount.
Recently there emerged much more striking problem because of the observation
of BH with ${M \sim 100M_\odot}$. Formation of such black holes 
in the process of stellar collapse was considered  to be strictly  forbidden.
On the  other hand, primordial black holes with the observed by LIGO masses may be easily created 
with sufficient density. \\
{\bf 2. Formation of BH binaries from the original stellar binaries. }
Stellar binaries are formed from common interstellar gas clouds and are quite frequent in galaxies.
If BH is created through stellar collapse,  a small non-sphericity would result in a huge 
velocity of the BH and the binary would be destroyed.
BH formation from PopIII stars and subsequent formation of BH
binaries with  tens of ${M_\odot}$ is estimated to be small.
{The problem of the binary formation is simply solved if the observed sources of GWs are the binaries of
primordial black holes.} 
{They were at rest in the comoving volume, and when inside the cosmological horizon they were gravitationally attracted and  
might loose energy due to dynamical friction or interaction with third body in the early universe.
The probability for them to become gravitationally binded is probably high enough.}
The conventional astrophysical scenario is not excluded but less natural.\\
{\bf 3.  Low spins of the coalescing BHs.}
{The low values of the BH spins sae observed
in GW150914 and in almost all (except for three) other events.}
It strongly constrains astrophysical BH formation from close binary systems. 
{Astrophysical BHs are expected to have considerable angular momentum, nevertheless the
dynamical formation of double massive low-spin BHs in dense stellar clusters is not excluded, though difficult.} 
{On the other hand, PBH practically do  not rotate, because vorticity perturbations 
in the early universe are vanishingly small.}
{Still, individual PBH forming a binary initially rotating on elliptic orbit could gain collinear spins 
about 0.1 - 0.3, rising with the PBH masses and eccentricity~\cite{Post-Mit,PKM}.
This result is in agreement with the
GW170729 LIGO event produced by the binary with masses ${50 M_\odot}$  and ${30 M_\odot}$ 
and  GW190521.

To summarize: each of the mentioned problems might be solved in the conventional frameworks but it looks
much simpler to assume that the LIGO/Virgo/Kagra sources are primordial black holes.
 

\subsection{Chirp mass distribution  \label{ss-chirp} }

{It is well known that two rotating gravitationally bound massive bodies emit gravitational 
waves.} In quasi-stationary inspiral regime, the radius of the orbit and the rotation frequency
are approximately constant, to be more exact, slowly decreasing, and the GW frequency is twice the rotation frequency. 
{The luminosity of the GW radiation in inspiral regime is:}
\be {{
L  = \frac{32}{5}\,m_{Pl}^2\left(\frac{M_c\,\omega_{orb}}{m_{Pl}^2}\right)^{10/3}\,,
}}
\ee
where $M_1$, $M_2$ are the masses of two bodies in the binary system and 
${M_c}$ is the so called chirp mass: 
\be {{
M_c=\frac{(M_1\,M_2)^{3/5}}{(M_1+M_2)^{1/5}} \, ,
}}
\ee
and 
\be {{
\omega^2_{orb} =  \frac{M_1+M_2}{m_{Pl}^2 R^3}\,.
} }
\ee

In ref.~\cite{dkmppss}
the available data on the chirp mass distribution of the black holes in the coalescing binaries in O1-O3 LIGO/Virgo runs are
analyzed and compared with theoretical expectations based on the hypothesis that these black holes are primordial with 
log-normal mass spectrum. The results are presented in \ref{fig-EDF}.
{The inferred best-fit mass spectrum parameters, }
{$M_0=17 M_\odot$ and ${\gamma=0.9}$,} fall 
within the theoretically expected range and shows excellent agreement with observations. 
{On the opposite, binary black hole formation based }
{on massive binary star evolution} require additional adjustments to 
reproduce the observed chirp mass distribution, see Fig. \ref{fig-EDF}.
Similar value of the parameters are 
obtained in refs.~\cite{Raidal,Liu}, see also~\cite{KAP-Mit}.

\begin{figure}[htbp]
\begin{center}
\includegraphics[scale=0.25]{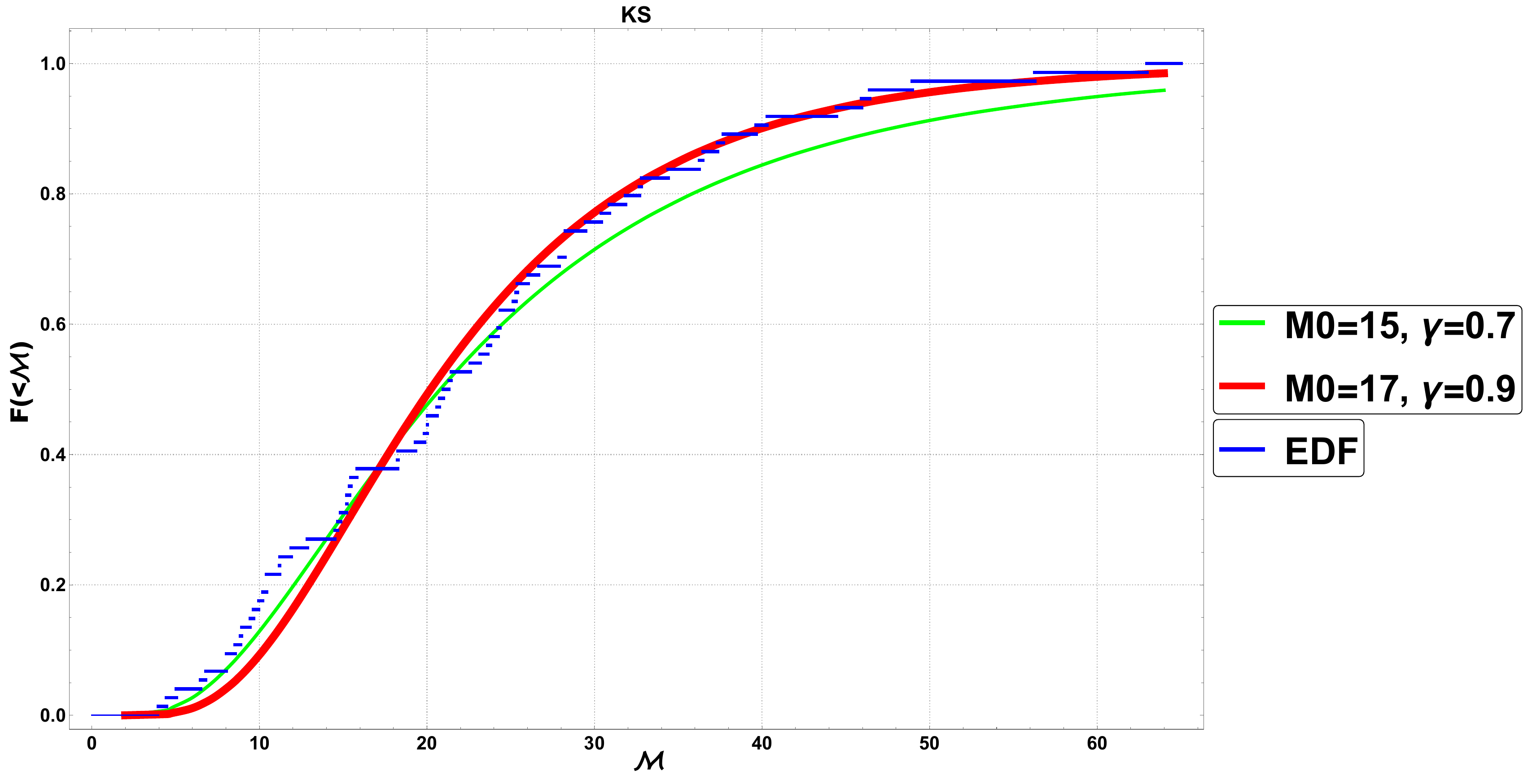}
\caption{Model distribution $F_{PBH}(< M)$ with parameters  $M_0\approx 17M_\odot$ and 
${\gamma\sim 1}$ for two best Kolmogorov-Smirnov tests.  EDF= empirical distribution function.}
\label{fig-EDF}
\end{center}
\end{figure}

\begin{figure}[htbp]
\begin{center}
\includegraphics[scale=0.3]{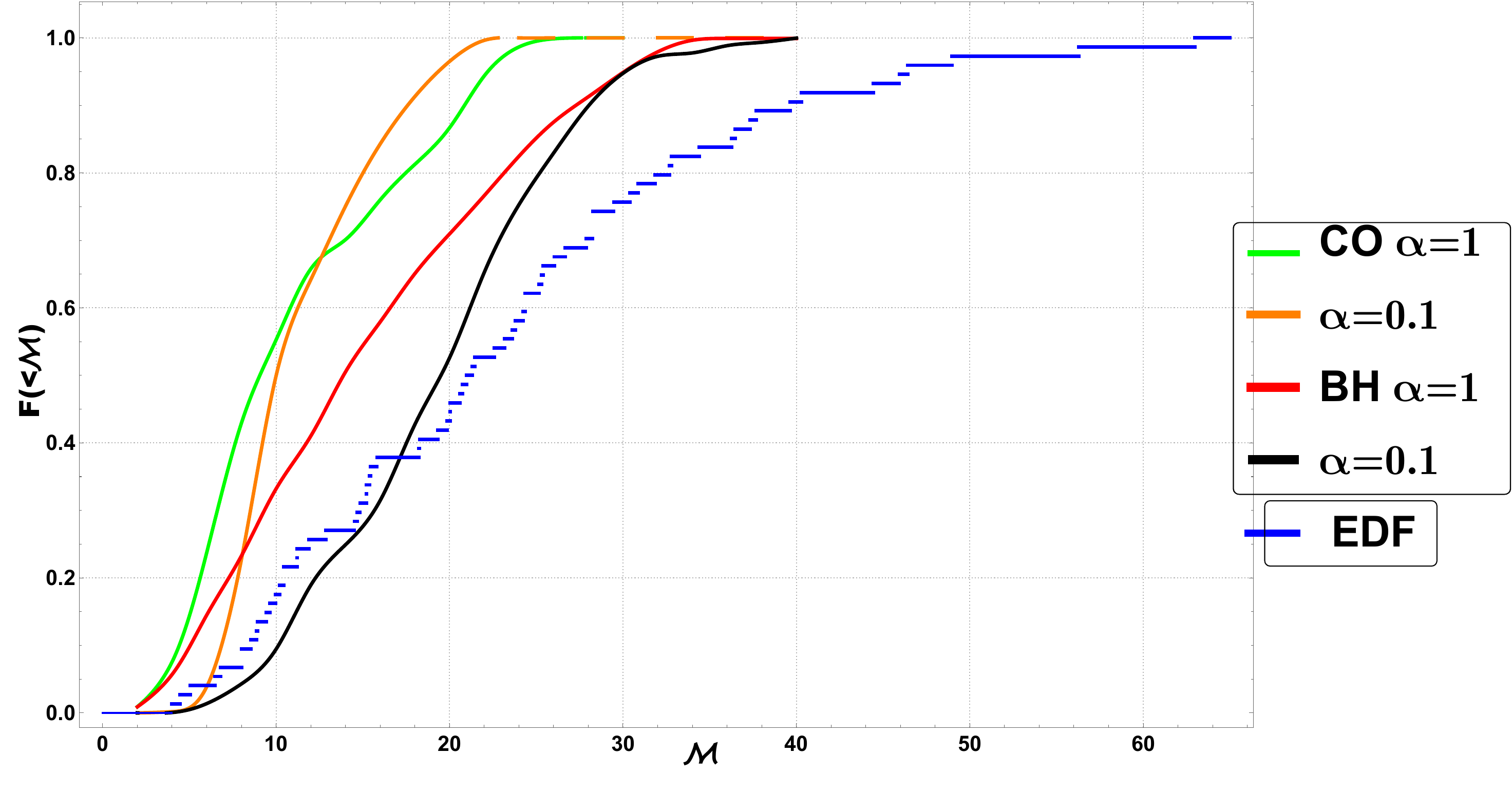}
\caption{Cumulative distributions $F(< M)$ for several {astrophysical} models of binary BH coalescences.}
\end{center}
\end{figure}

So we may conclude that
{ PBHs with log-normal mass spectrum perfectly fit the data. }
{Astrophysical black holes seem to be disfavoured.}

 
{A new analysis of the Ligo-Virgo-Kagra data was performed recently in ref.~\cite{KAP-Mit}, see fig.~\ref{f-kap}.
 The authors concluded that the chirp-mass distribution of LVK GWTC-3 BH+BH binaries with distinct two bumps can be explained by two different populations of BH+BH binaries: \\
1) the low-mass bump at $M_0 \sim 10 M_\odot$ due to the astrophysical BH+BH formed in the local Universe from the evolution of massive binaries \\
2) the PBH binaries with log-normal mass spectrum  with 
$M_0\simeq 10 M_\odot$ and $\gamma\simeq 10$. The central mass of the PBH distribution 
is larger than the expected PBH mass at the QCD phase transition ($\sim 8 M_\odot$) but still can be 
accommodated with the mass of the cosmological horizon provided that the temperature 
$T_{QCD}\sim 70$~MeV, possible for non-zero chemical potential at QCD p.t.
The model includes almost equal contributions from coalescences of astrophysical binary BHs 
(green dashed curve) and primordial BHs with the initial log-normal mass spectrum 
with parameters $M_0=33 M_\odot$, $\gamma=10$ - with such $\gamma$ heavier PBH practically are 
not created.


\begin{figure}[htbp]
\begin{center}
\includegraphics[scale=0.30]{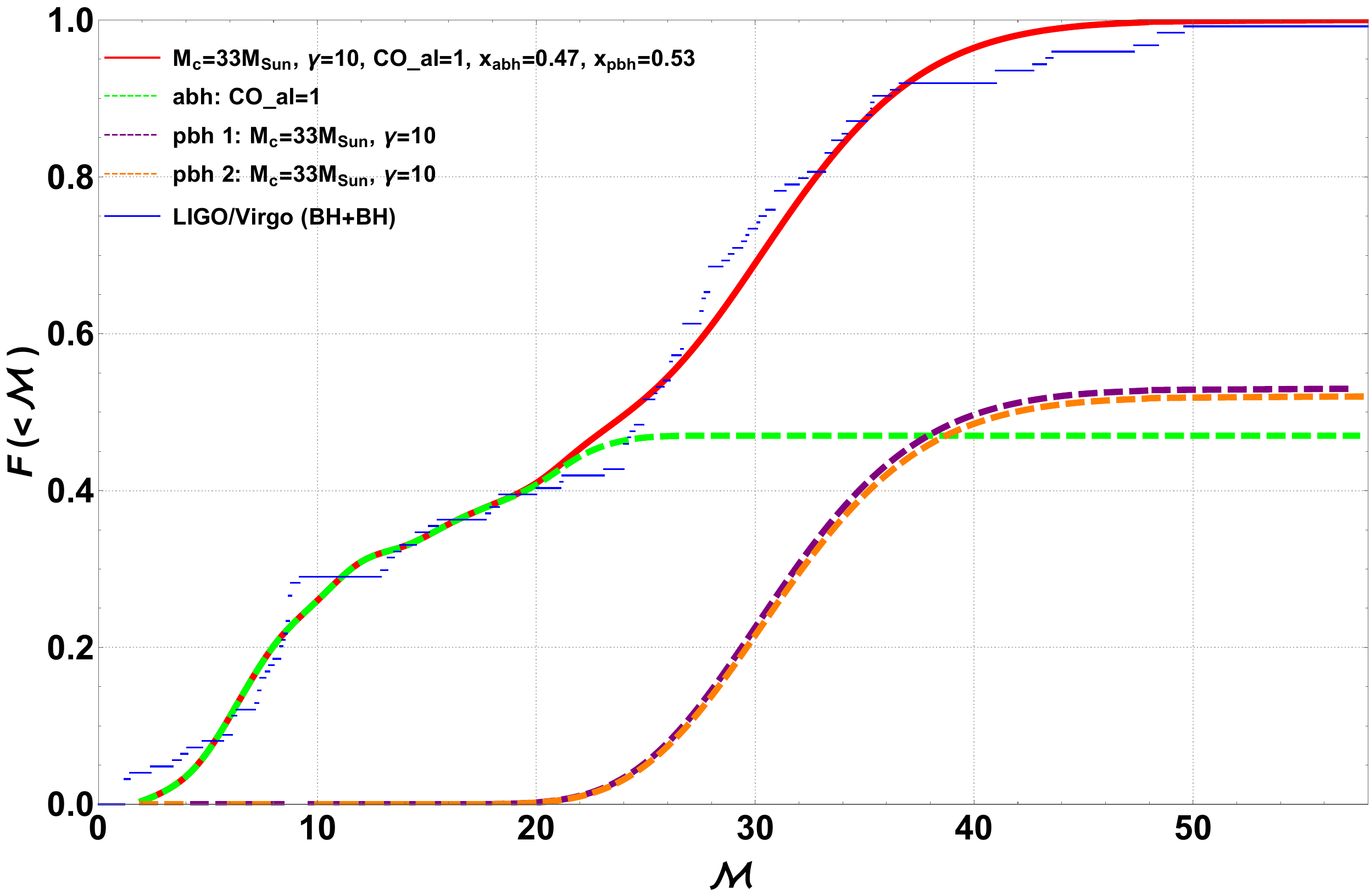}
\caption{The observed (blue step-like curve) and model (red solid curve) distribution function of the chirp-masses of 
coalescing binary BHs from the LVK GWTC-3 catalogue. 
(green dashed curve) and primordial BHs with the initial log-normal mass spectrum 
with parameters $M_0=33 M_\odot$, $\gamma=10$.} 
\vspace{-3mm}
\label{f-kap}
\end{center}
\end{figure}



 \section{Early universe conundrums \label{s-early-problem}}
  
During last decade observations made by Hubble Space Telescope (HST),
see e.g. ~\cite{hst-1,hst-2,hst-3}
and very recently by James Webb Space Telescope (JWST)~\cite{jwst-1,jwst-2,jwst-3,jwst-4}
 have led to the surprising conclusion  that the early universe, younger 
than one billion years, is densely populated by well developed galaxies, quasars (supermassive
 black holes), gamma-bursters, and heavy elements (heavier than helium).
 These striking results were taken by the community as absolutely incompatible with the canonical 
$\Lambda$CDM cosmology, especially after release of the JWST data. In fact already observations
of HST could be a sufficient cause for anxiety, not only with respect to the early universe but also to the 
contemporary very old universe almost 15 billion years old. The troubling situation in the
present day universe as well  as in the universe with redshifts $z= 6-10$ are 
summarised in review~\cite{AD-UFN}. 
The state of art is emphatically characterised as crisis in cosmology that 
is believed to hit a strong blow to the conventional $\Lambda$CDM picture.

However, the resolution of the above mentioned problems was suggested 
in our papers~\cite{DS} (DS) and \cite{DKK}  (DKK). 
long before these problems arose. In these works a new
mechanism of massive primordial black hole (PBH) formation was worked out
that could lead to their efficient creation with the masses in the range from a fraction of the solar mass 
up to billion solar masses. 

An essential input of  DS and DKK papers is the suggestion of 
an inverted formation mechanism of galaxies and their central black holes.
Usually it is assumed that supermassive BHs (SMBHs), that are
observed in centres of all large galaxies, are created by matter accretion
to the density excess in the galactic centre, 
but the estimated necessary time is much longer than the universe age, 
even for the contemporary universe, with the age about 15 billion years,
to say nothing about the 20 times younger universe at $z\sim 10$. 

On the opposite, as it was  conjectured in refs.~\cite{DS,DKK}, supermassive black holes were created first in the 
early universe at prestellar epoch, that's why they are caller primordial, and later they have SEEDED galaxy formation.  

Model DS/DKK is verified by a very good agreement of the calculated log-normal
mass spectrum of PBH with observations and by 
discovery of abundant antimatter population in the Galaxy envisaged according to  DS and DKK.
The model also predicts an early formation of galaxies, seeded by PBH,
quasars (alias SMBH), rich chemistry (heavy elements), and dust in the early universe.


 \section{Cosmic antimatter \label{s-ant}}
 
 \subsection{Anti-history \label{ss-history}}
 
The father of antimatter is justly admitted to be Paul Dirac.
In his Nobel Lecture at December 12, 1933 
 ``Theory of electrons and positrons'', dedicated to his prediction of positrons,  he foresaw that there could be 
antistars and possibly antiworlds: ''... It is quite possible that... these stars being built up mainly of positrons 
and negative protons. In fact, there may be half the stars of each kind.
The two kinds of stars would both show exactly the same spectra, 
and there would be no way of distinguishing them by present astronomical methods.'' 

Now we expect that there may be some presumably small fraction of antistars in the Galaxy. 
Since they are immersed into interstellar
gas consisting of matter, they can be detected due to excess of  gamma radiation with energy of a several 
hundred MeV, originating from annihilation of the interstellar gas on the surface of antistar.

Situation is different if an antistar "lives" in a distant antigalaxy. Still it is possible in principle  to distinguish a star from an antistar through
rather subtle effects considered in ref.~\cite{DVN}. First of all
the spectra of the emitted radiation are not exactly the same,} {even if CPT is unbroken} 
and the polarization of radiation from weak decays
could be a good indicator and lastly the types of emitted neutrinos versus antineutrinos from supernovae or antisupernovae.

It is in fact surprising that Dirac was not the first person to talk about antimatter.
 In 1898, 30 years before Dirac and one year after discovery
of electron (J.J. Thomson, 1897)
Arthur Schuster~\cite{shuster} 
(another British physicist) conjectured that 
there might be other sign electricity, {\bf antimatter}, and 
supposed that there might be 
entire solar systems, made of antimatter, indistinguishable from ours.
Schuster made fantastic  wild guess that matter and antimatter are 
capable to annihilate and produce vast energy.

Schuster believed that they were gravitationally 
repulsive having negative mass. Two such objects on close contact
should have vanishing mass!?

Quoting paper~\cite{shuster}: ``When the year's work is over and all sense of responsibility has 
left us, who has not occasionally set his fancy free to dream about 
the unknown, perhaps the unknowable?'... Astronomy, the oldest and yet most
juvenile of the sciences, may still have some surprises in
store. May antimatter be commended to its case".

According to the classical scenario of the generation of the cosmological baryon asymmetry, proposed by A.D. Sakharov~\cite{sakharov} ,
 the baryon excess in the universe is on the average homogeneous, having the same sign determined by the sign of symmetry breaking 
 between particles and antiparticles. However, there are plenty of mechanisms leading to space varying amplitude of C and CP violations
 with possible sigh changing. If this is the case, the sign of baryon asymmetry could also be different leading to formation of matter and
 antimatter domains in the universe. Possible models of C and CP violation in cosmology that possess  this property
 are reviewed in~\cite{AD-CP}.

 \subsection{Matter and antimatter in the Universe \label{anti-universe}}
 
 To the best of my knowledge the pioneering  papers on cosmological antimatter were published 
 by F. Stecker in 1971~\cite{stecker-1}. But first suggestion that there might exist antimatter in the Galaxy
 were put forward in
 three year earlier in papers by Konstantinov et al~\cite{konstan-1,konstan-2} on search of antimatter in the Galaxy, see below 
 subsection \ref{ss-anti-galaxy}. Further development of the idea of matter-antimatter domain structure of the universe was presented
 in ref.~\cite{stecker-2}.
 
The analysis, performed  in ref.~\cite{C-AdR-G}, permits to conclude that matter–antimatter symmetric universe 
or close to that, is excluded by the observed
 cosmic diffuse gamma-ray background and a distortion of the cosmic microwave background.
 
However, there still remains some space for the fraction of cosmological antimatter, though
considerably restricted. It is argued by G. Steigman
in ref.~\cite{GS-1} that the nearest anti-galaxy should be out of our galaxy cluster and thus could not
be closer than at $\sim10$ Mpc.
In a subsequent paper by the same author~\cite{GS-2} it is argued that
the fraction of antimatter in Bullet Cluster should be below  $ 3\times 10^{-6}$. 

 Summary of the situation of the year 2002 was presented  in two keynote lectures at 14th  
Rencontres de Blois on Matter - Anti-matter Asymmetry~\cite{stecker-Blois,dolgov-Blois}.

  \subsection{Antimatter in Milky Way \label{ss-anti-galaxy}}
 
The search of antimatter in the Milky Way was intiiated 
 by Konstantinov with coworkers in 1968~\cite{konstan-1,konstan-2}.
 This activity was strongly criticised by Ya.B.  Zeldovich despite very friendly relations between the two.
 In agreement with canonic faith no antimatter may exist in our Galaxy and that explains negative attitude
 of Zeldovich to Konstantinov activity. Until recently there was no reason to suspect that any noticeable
amount of antimatter might be in the Galaxy. The predictions of refs.~\cite{DS,DKK} were not taken seriously.
Now there are  a lot of data indicating that Milky Way contains siginificant amount 
of antimatter of different kinds: positrons, antinuclei, 
and possibly antistars.  The observations do not violate the existing  bounds on galactic antimatter. 
According to the predictions of papers~\cite{DS,DKK} antimatter objects could be not only in the Galaxy
but in its halo as well.

According to ref.~\cite{ballmoos-bound}, the analysis of the intensity of gamma rays created by the Bondi
accretion of interstellar gas to the surface of an antistar would 
allows to put a limit on the relative density of antistars in the Solar neighbourhood: $N_{\bar *} /N_* < 4 \cdot 10^{-5} $
inside 150 pc from the Sun.

The bounds on galactic antimatter are analysed in refs.~\cite{bd,db,bdp}. 
The limits on the density of galactic antistars are rather loose, because the annihilation proceeds 
only on the surface of antistars, i.e on the objects with short mean free path of protons, so the extra luminosity
created by matter-antimatter annihilation is relatively low. \\[1mm]
{\bf Anti-evidence: cosmic positrons.} \\
Existence of rich populations of positrons in the Galaxy was noticed long ago through the observations of 511 keV gamma ray 
line (see~\cite{anti-e1, anti-e2, anti-e3} and references therein) with the flux 
\be\label{flux} {{
\Phi_{511 \; {\rm keV}} = { 1.07 \pm 0.03 \cdot 10^{-3} }\; 
{\rm photons \; cm^{-2} \, s^{-1}} .
}}
\ee
The width of the line is about 3 keV. 
The emission mostly goes from the  Galactic bulge and  at much lower level from the disk.
This unambiguously indicates the frequent annihilation of nonrelativistic $e^+ e^-$ pairs in the Galactic bulge with the rate~\cite{anti-e1}
\be \label{Neebulge}
\dot N_{ee}^\text{bulge} \sim 10^{43} \text{ s}^{-1}.
\ee

Note that one of the brightest X-ray  sources in the region around the Galactic Center   
got the name Great Annihilator~\cite{gr-ann}. Possibly it is a microquasar first detected in soft X-rays by the Einstein 
Observatory~\cite{Ein-observ} and later detected in hard X-rays by the space observatory ``Granat''~\cite{granat}.

There is no commonly accepted point of view on the origin of the cosmic positrons. The conventional hypothesis that positrons are created in strong 
magnetic fields of pulsars is at odds with the AMS data~\cite{AMS-19}. 
{However, this conclusion  is questioned in ref.~\cite{kachel} where it is shown that
{these features could be consistently explained by a nearby source which was active 
$\sim 2$ Myr ago and has injected  $(1-2)\times10^{50}$ erg in cosmic rays.

A competing option is that positrons are created by the Schwinger process at the horizon of small black holes with masses  $\gtrsim10^{20}$ g . 
This mechanism was suggested in ref.~\cite{BDP} and discussed in more detail in ref.~\cite{AD-AR}.

One more possibility that is closer to the spirit of this talk
is that positrons are primordial, produced in the early universe in relatively small antimatter domains~\cite{DS, DKK}. 
Possible observation of the unexpectedly high flux of antinuclei~\cite{anti-nuc-AMS-1, anti-nuc-AMS-2} and antistars in the Galaxy~\cite{anti-stars} 
strongly supports this hypothesis, in particular, in ref.~\cite{,bbbdp,bpbbd},  it is advocated that
antihelium cosmic rays are created by antistars.\\[1mm]
{\bf Anti-evidence: cosmic antinuclei.} 

In 2018 AMS-02 announced possible observation of six
${\overline{He}^3}$ and two ${\overline{He}^4}$~\cite{choutko,ting-CERN}.
Accumulated by 2022 data contains some more events:
7 $\overline D$ (at energies $\lesssim 15$ GeV) and 9 $\overline{He}^4$ at ($E\sim 50$ GeV).
These numbers correspond roughly speaking  to ${\overline{He}/He \sim10^{-9}}$.
This number is much larger than the expected number of
$\overline{He}^4$, if it were created in cosmic ray collisions.
It is possible that the total flux of anti-helium is even much higher because low energy 
${\overline{He}}$ may escape registration in AMS.

The probability of the secondary creation of different antinuclei was estimated in ref.~\cite{cosm-anti-nuc}.
As shown in this  paper,  anti-deuterium could be  
most efficiently produced in the collisions ${\bar p\,p}$ or
${\bar p\, He}$ that can create the flux
${\sim 10^{-7} /m^{2}/ s^{-1}} $/steradian/GeV/neutron),
i.e. 5 orders of magnitude below the observed flux of antiprotons.
Antihelium could be created in the similar reactions and 
the fluxes of  ${\overline{He}^3}$ and ${\overline{He}^4}$, that
could be created in 
cosmic rays would respectively be 4 and 8
orders of magnitude smaller than the flux of the secondary created anti-D.

According to the works~\cite{DS,DKK}, antinuclei should be
primordial i.e. created in the very early universe during  
big bang nucleosynthesis (BBN) inside antimatter bubbles with high baryon density. 
However, the standard anti-BBN surely does not help, since normally BBN gives 75\% of hydrogen, 25\% of helium-4,
and a minor fraction of deuterium, at the level a few times $10^{-5}$,
in a huge contrast to the observed ratio of anti-deuterium to anti-helium which is of order unity. 
The same problem exists for the ratio
of $\overline{He}^3$ to $\overline{He}^4$, that is also of order unity instead of the standard  $\sim 3\times10^{-5}$.

If we assume that 
in the model of~\cite{DS,DKK}  the abundances of anti-D and anti-He are determined by normal BBN with large baryon-to-photon
ratio $\beta\sim1$, the problem would be even more pronounced, because amount of deuterium and helium-3 would be negligibly small,
even much less than $10^{-5}$.
On the other hand in our scenario  formation of primordial elements takes place inside non-expanding compact 
stellar-like objects with practically fixed temperature. If the temperature is sufficiently high, this so called BBN may stop 
with almost equal abundances of D and He. One can see that looking at 
abundances of light elements at a function of temperature. 
{If it is so, antistars may have equal amount of $\overline{D}$ and $\overline{He}$\\[1mm]
{\bf Anti-evidence: antistars in the Galaxy.}\\
A striking announcement of the 
possible discovery of anti-stars in the Galaxy was made in ref.~\cite{antistars}
The catalog 14 antistar candidates was identifyed, not associated with any objects belonging 
to established gamma-ray source classes and with a spectrum compatible with baryon-antibaryon annihilation.
There results are illustrated in Fig.~\ref{fig:sources}.

\begin{figure}[htbp]
\includegraphics[scale=0.85]{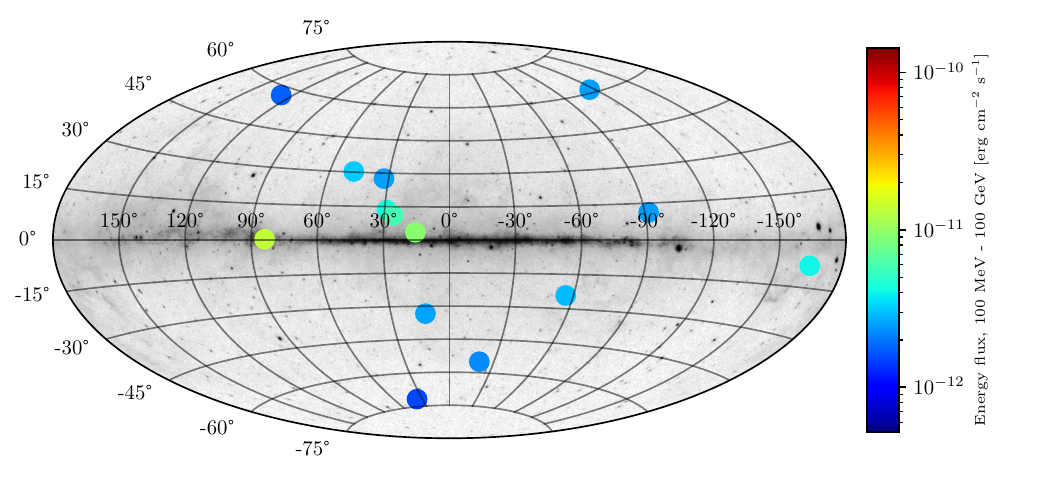}
\caption{\label{fig:sources} Positions and energy flux in the 100 MeV - 100 GeV range of antistar candidates selected in 4FGL-DR2. 
Galactic coordinates. The background image shows the Fermi 5-year all-sky photon counts above 1 GeV 
}
\end{figure}  


In ref.~\cite{bbbdp} a supplimentary method of antistar identification was proposed.
In astrophysically plausible cases of the interaction of neutral atmospheres or winds from 
antistars with ionised interstellar gas, the hadronic annihilation 
{ will be preceded by the formation of excited ${p \bar p}$
and $He {\bar p}$ atoms.} 
These atoms rapidly cascade down to low levels prior to 
annihilation giving rise to a series of narrow lines which can be associated with the hadronic 
annihilation gamma-ray emission. The most significant are L (3p-2p) 1.73 keV line (yield more 
than 90\%) from ${p \bar p}$ atoms, and M (4-3) 4.86 keV (yield $\sim 60$\%) and L (3-2) 11.13 
keV (yield about 25\%) lines from $He^4 \bar p$ atoms. These lines can be probed in dedicated 
observations by forthcoming sensitive X-ray spectroscopic missions XRISM and Athena and in 
wide-field X-ray surveys like SRG/eROSITA all-sky survey.

Possible impact of antistars on Big Bang Nucleosynthesis are discussed in ref.~\cite{anti-BBN},
based on consideration of  the measurement of cosmic
ray antihelium nuclei by AMS-02 in ref.~\cite{anti-bbn}. 

Connection between antistars and the observed fluxes of antinuclei was studied in recent paper ~\cite{bpbbd}.
A minor population of antistars in galaxies has been predicted by some of non-standard
models of baryogenesis and nucleosynthesis in the early Universe, and their presence is not yet excluded
by the currently available observations. Detection of an unusually high abundance of antinuclei in
cosmic rays can probe the baryogenesis scenarios in the early Universe. 

It is shown~\cite{bpbbd}  that the flux of antihelium cosmic rays
reported by the AMS-02 experiment can be explained by 
{Galactic anti-nova outbursts, thermonuclear
anti-SN Ia explosions, a collection of flaring antistars, or an extragalactic source} with abundances not
violating existing gamma-ray and microlensing constraints on the antistar population.


\section{Mechanism of creation of PBH and antimatter  \label{s-creation}}

The mechanism of PBH and galactic antimatter creation~\cite{DS,DKK} is essentially based on the scenario of
supersymmetry (SUSY) motivated baryogenesis, proposed  by Affleck and Dine (AD)~\cite{AD}.
SUSY generically predicts existence of  scalars, $\chi$, with non-zero baryonic number,
${ B\neq 0}$. Another prediction of high energy SUSY models is an existence of flat directions in the 
$\chi$-potential, either quartic (self-interaction): 
\be{{
U_\lambda(\chi) = \lambda |\chi|^4 \left( 1- \cos 4\theta \right)
}}
\ee
or quadratic, i.e. the mass term, ${ U_m=m^2 \chi^2 + m^{*\,2}\chi^{*\,2}}$:
\be
U_m( \chi ) = m^2 |\chi|^2} {\left[{ 1-\cos (2\theta+2\alpha)} 
\right],
\ee
where ${ \chi = |\chi| \exp (i\theta)}$ and ${ m=|m|e^\alpha}$. {If ${\alpha \neq 0}$, C and CP are  broken.} Potential energy 
does not rise along these flat directions.

It is shown in several papers that 
at the inflationary epoch the average value of $\chi^2$ linearly rises with time~\cite{vil,adl,aas}, see also~\cite{ad-dp}, 
asymptotically reaching the value $\langle \chi^2 \rangle = 3H^4/(8 \pi^2 m_\chi^2)$, where $m_\chi$ is the mass of the 
scalar $\chi$ and $H$ is the Hubble parameter in the course of inlation. Due to that $\chi$ bosons may condense along flat directions 
of the quartic potential, when and if its mass was smaller
than the inflationary Hubble parameter.  In papers~\cite{vil,adl,aas}, the result is obtained by (questionable?) infrared regularisation,
while in ref~\cite{ad-dp} rigorous equation of motion for $\langle \chi^2 \rangle $ was applied. All the works agree with the absolute asymptotic value 
of $\langle \phi^2 \rangle$ and with 
the law of rising with time $\langle \chi^2 \rangle = (8 \rho t)/(3H)$, where $\rho$ is the cosmological energy density in the course
of inflation. However the signs are opposite with $\langle \chi^2 \rangle = -(8 \rho t)/(3H)$ in ref.~\cite{ad-dp}

Note that in GUT SUSY baryonic number is naturally non-conserved, because of generic non-invariance of ${U(\chi)}$
w.r.t. phase rotation $\chi \rightarrow \chi \exp( i\theta)$.  

After inflation ${\chi}$ was far away from origin due to rising quantum fluctuations and, when 
inflation ends, it started to evolve down to the equilibrium point, ${\chi =0}$,
according to equation of motion that formally coincides with the equation of motion of a point-like particle in Newtonian mechanics:
\be{{
\ddot \chi +3H\dot \chi +U' (\chi) = 0.
}}
\ee
The baryonic number  of $\chi$:
\be{{
B_\chi =\dot\theta |\chi|^2
}}
\ee
is analogous to mechanical angular momentum in complex plane  $[Re \chi, Im \chi]$. 
After ${{\chi}}$ decays, the accumulated baryonic number of $\chi$ is transferred into
baryonic number of quarks in B-conserving process.
AD baryogenesis could lead to baryon asymmetry of order of unity, much larger
than the observed  asymmetry ${\sim10^{-9}}$.

If $ { m\neq 0}$ and the flat direction of quadratic and quartic valleys are different,
the angular momentum, B, would be generated by the "rotation" induced by the motion of $\chi$ from quartic flat direction to the
quadratic one. In other words, the field $\chi$ would acquire non-zero baryonic number, generically very large .

{If CP-odd phase ${\alpha}$ is non-vanishing, both baryonic and 
antibaryonic domains might be  formed}
{with possible dominance of either of them.}
{Matter and antimatter objects may exist but possibly global ${ B\neq 0}$.}
 
An essential development proposed in works~\cite{DS,DKK} was an introduction of the
new interaction between the Affleck-Dine field and the inflaton
$\Phi$, the first term in the equation below:
\be 
U = {g|\chi|^2 (\Phi -\Phi_1)^2}  +
\lambda |\chi|^4 \,\ln \left( \frac{|\chi|^2 }{\sigma^2 } \right)
+\lambda_1 (\chi^4 + h.c. ) + 
(m^2 \chi^2 + h.c.). \,\,\,\,\,\,\,\,\,\,\,\,\,\,\,\,\,\,\,\,
\ee
This coupling between $\chi$ and the inflaton is the general renormalizable interaction of two scalar field.
The only tuning is the assumption that $\Phi$ reaches the value $\Phi_1$ during inflation significantly before it ends, with the remaining 
number of e-foldings about 30-40.

The window to the flat directions is open, near ${\Phi = \Phi_1}$. At that period
the field ${\chi}$ could rise to large values, according to the quantum diffusion
equation derived by Starobinsky, generalised to a complex field $\chi$.

If the window to flat direction, when ${\Phi \approx \Phi_1}$ is open only {during 
a short period,} cosmologically small but possibly astronomically large 
bubbles with high baryon-to-photon ratio ${\beta}$ could be
created, occupying a tiny
fraction of the total universe volume, while the rest of the universe has the observed
$\beta \approx 6\cdot 10^{-10}$, created 
by the normal small $\chi$. 
{The fundament of PBH creation has been build at inflation by making large isocurvature
fluctuations at relatively small scales, with practically vanishing density perturbations.} 

The initial isocurvature perturbations are contained in large baryonic number of massless quarks in rather small bubbles. We
call them high baryonic bubbles, HBBs.
Density perturbations were generated rather late after QCD phase transition, at temperatures around 100 MeV,
when massless quarks turned into massive baryons. The resulting high density contrast could lead to creation of PBHs.
The mechanism is very much different from any other described in the literature models of PBH formation.

{The emerging universe looks like a piece of Swiss cheese, where holes are high baryonic 
density objects occupying a minor fraction of the universe volume.}\\[1mm]

Outcome of the DS/DKK mechanism:
\begin{itemize}
\item
PBHs with log-normal mass spectrum - confirmed by the observations!
\item
{Compact stellar-like objects, similar to cores of red giants.}
\item
{Disperse hydrogen and helium clouds  with (much) higher than average ${n_B}$ density.} 
\item
Strange stars with unusual chemistry and velocity.
\item
{${\beta}$ may be negative leading to creation of
(compact?) antistars which could survive annihilation despite being submerged into the 
homogeneous baryonic background.}
\item
Extremely old stars could exist  and indeed they are observed, even,
"older than universe star"  is found; its prehistoric age is mimicked by the unusual initial chemistry.  

\end{itemize}

{The mechanism of PBH creation pretty well verified by the data on the BH mass spectrum and 
on existence of antimatter in the Galaxy, especially of antistars. So we may expect that it indeed solves
the problems created by HST and JWST.}

Thus we may conclude that canonical  $\Lambda$CDM cosmology is saved by PBHs.}
{Antimatter in our backyard is predicted and found.}



\begin{thebibliography}{99}

\bibitem{hst-1}
A. Monna, S. Seitz, N. Greisel, {\it et al} 
"CLASH: $z \sim 6$ young galaxy candidate quintuply lensed by the frontier field cluster RXC J2248.7-4431"
arXiv:1308.6280v2 [astro-ph.CO] 6 Dec 2013

\bibitem{hst-2}
W. Zheng, A. Zitrin, L. Infante, 
"Young Galaxy Candidates in the Hubble Frontier Fields IV. MACS J1149.5+2223",
arXiv:1701.08484 [astro-ph.GA].

\bibitem{hst-3}
P.A. Oesch, G. Brammer, P.G. van Dokkum, {\it et al},
"A Remarkably Luminous Galaxy at z=11.1 Measured with Hubble Space Telescope Grism Spectroscopy",
arXiv:1603.00461 [astro-ph.GA].

\bibitem{jwst-1}
S.L. Finkelstein, M.B. Bagley, H.C. Ferguson, S.M.
Wilkins, J. S. Kartaltepe, et al., "CEERS Key Paper I: 
An Early Look into the First 500 Myr of Galaxy Formation with JWST",
Astrophys. J. Lett. 946,
L13 (2023), arXiv:2211.05792 [astro-ph.GA].

\bibitem{jwst-2} 
 Y. Harikane, M. Ouchi, M. Oguri, Y. Ono, K. Nakajima, et al., 
 "A Comprehensive Study on Galaxies at z~9-16 Found in the Early JWST Data: UV Luminosity Functions and 
 Cosmic Star-Formation History at the Pre-Reionization Epoch",
 Astrophys. J., Suppl. Ser. 265, 5 (2023), arXiv:2208.01612 [astro-ph.GA].

\bibitem{jwst-3}
M. Castellano, A. Fontana, T. Treu, P. Santini,
E. Merlin, et al., 
Early results from GLASS-JWST. III: Galaxy candidates at $z \sim 9-15$
Astrophys. J. Lett. 938, L15 (2022),
arXiv:2207.09436 [astro-ph.GA].

\bibitem{jwst-4}
P. Santini, A. Fontana, M. Castellano, N. Leethochawalit, M. Trenti, et al., 
Early results from GLASS-JWST. XI: Stellar masses and mass-to-light ratio of $z>7$ galaxies
Astrophys. J. Lett. 942, L27 (2023), arXiv:2207.11379 [astro-ph.GA].

\bibitem{AD-UFN}
A.D. Dolgov,  "Massive and supermassive black holes in the contemporary and early Universe and problems in 
cosmology and astrophysics",  Phys. Usp. 61 (2018) 2, 115.

\bibitem{DS}
A.Dolgov,  J.Silk, 
{"Baryon isocurvature fluctuations at small scale and baryonic  dark matter}"
PRD 47 (1993) 4244.

\bibitem{DKK}
A.Dolgov, M.Kawasaki,  N.Kevlishvili, 
"Inhomogeneous baryogenesis, {cosmic antimatter,} and dark matter",
Nucl. Phys. B807 (2009) 229.

\bibitem{spectrum-z11}
S. Tacchella, et al  ''JADES Imaging of GN-z11: Revealing the Morphology and Environment of a
 Luminous Galaxy 430 Myr After the Big Bang",
arXiv:2302.07234

\bibitem{spectrum-z11-2}
A.J. Bunker, { et al}  
"JADES NIRSpec Spectroscopy of GN-z11: Lyman-$\alpha$ emission and
  possible enhanced {nitrogen} abundance in a $z=10.60$ luminous galaxy,"
arXiv:2302.07256.

\bibitem{alma-z12}
 T. Bakx, et al, 
'' Deep ALMA redshift search of a z~12 GLASS-JWST galaxy candidate'',
 arXiv:2208.13642 

\bibitem{imp-gal}
I. Labbé
"A population of red candidate massive galaxies 600 Myr after the Big Bang'',
Nature, Volume 616, Issue 7956, p.266, arXiv:2207.12446

\bibitem{alma-6.8}
R. Endsley et al,
''ALMA confirmation of an obscured hyperluminous radio-loud AGN at z = 6.853 associated with a dusty starburst in the 1.5 deg2 COSMOS field'', 
MNRAS, {\bf 520} Issue 3, 2023, Pages 4609–4620

\bibitem{peng-metals}
B. Peng, et al,  
"Discovery of a Dusty, Chemically Mature Companion to a $z \sim 4$ Starburst Galaxy in JWST ERS Data'',
The Astrophysical Journal Letters, {\bf  944}, 2023, Issue 2, id.L36.

\bibitem{cameron}
A.J. Cameron, et al, 
"Nitrogen enhancements 440 Myr after the Big Bang: super-solar N/O, a
  tidal disruption event or a dense stellar cluster in GN-z11?"
arXiv:2302.10142.

\bibitem{liu}
B. Liu, V. Bromm,
"Accelerating early galaxy formation with primordial black holes'',
 arXiv:2208.13178, accepted for publication in ApJL.

\bibitem{bogdan-1}
A. Bogdan, {\it et al}, 
''Detection of an X-ray quasar in a gravitationally-lensed z=10.3 galaxy suggests that early supermassive black 
holes originate from heavy seeds'',
arXive 2305.15458

\bibitem{seed-2}
A.D. Goulding, 
{\it et al},
'UNCOVER: The growth of the first massive black holes from JWST/NIRSpec -- spectroscopic confirmation of an X-ray luminous AGN at z=10.1''  arXiv:2308.02750.

\bibitem{tri-BH}
"The Galaxy: A Candidate Direct-collapse Supermassive Black Hole between Two Massive, Ringed Nuclei",
P. van Dokkum, G. Brammer, J. F. W. Baggen, {\it et al}
The Astrophysical Journal Letters, Volume 988, Number 1, July 15, 2025.

\bibitem{AD-KP-dwarfs}
A. Dolgov, K. Postnov, "Globular Cluster Seeding by Primordial 
Black Hole Population",  JCAP 04 (2017) 036, e-Print: 1702.07621 [astro-ph.CO]. 

\bibitem{seed-dwarf-1}
M. Mi{\'c}i{\'c}, et al,
''Two Candidates for Dual AGN in Dwarf-Dwarf Galaxy Mergers,''
arXiv:2211.04609 [astro-ph.GA]; accepted for publication in ApJ.

\bibitem{dwarf-2}
J. Yang et al, 
''Intermediate-mass black holes: finding of episodic, large-scale and powerful jet activity in a dwarf galaxy'',
Monthly Notices of the Royal Astronomical Society, Vol. 520, p. 5964 (2023)
e-Print: 2302.06214 [astro-ph.GA,astro-ph.HE].


\bibitem{prim-stars}
 J. Ying, {\it et al}, "The Absolute Age of M92",   
 Ap.J. 166 (2023) 1, 18,  e-Print: 2306.02180 [astro-ph.SR] 

\bibitem{pristine}
A. Arentsen, the University of Cambridge, presented 
at the National Astronomy Meeting 2023 at the University of Cardiff.

\bibitem{old-star-1}
Cowan J.J.,  { et al},
"The chemical composition and age of the metal-poor halo star BD+17$^o$ 3248"
{\it Astrophys. J.}  {\bf 572} (2002), 861 , astro-ph/0202429.

\bibitem{halo-age}
J. Kalirai 
''The Age of the Milky Way Inner Halo'',
 Nature 486 (2012) 90, e-Print: 1205.6802 [astro-ph.GA]

\bibitem{old-star-2}
A. Frebel, { et al.},
"Discovery of HE 1523-0901: A Strongly r-Process Enhanced Metal-Poor Star with Detected Uranium",
{\it Astrophys.J.}, {\bf 660} (2007), L117; astro-ph/0703414. 

\bibitem{older-universe} 
H.E. Bond, {\it et al},
''HD 140283: A Star in the Solar Neighborhood that Formed Shortly After the Big Bang".
{\it Astrophys. J. Lett.}, {\bf 765} (2013), L12. L12; arXiv:1302.3180.

\bibitem{vennes}
S. Vennes S, et al, 
"An unusual white dwarf star may be a surviving remnant of a subluminous Type Ia supernova'''
Science,  {\bf 357}, 2017, Issue 6352, p. 680.

\bibitem{hattori}
K. Hattori, et al., 
''Old, Metal-Poor Extreme Velocity Stars in the Solar Neighborhood'',
Astrophys.J. 866 (2018) 2, 121, e-Print: 1805.03194 

\bibitem{marchetti}
T. Marchetti et al
''Gaia DR2 in 6D: Searching for the fastest stars in the Galaxy'',
arXiv:1804.10607

\bibitem{ruffini}
N.J. Ruffini, A.R. Casey,
''A hyper-runaway white dwarf in Gaia DR2 as a Type Iax supernova primary remnant candidate'',
arXiv:1908.00670.

\bibitem{bennet}
D.P. Bennet et al,
''A Planetary Microlensing Event with an Unusually Red Source Star: MOA-2011-BLG-291'',
arXiv:1806.06106 [astro-ph.EP].

\bibitem{puls-hum}
G. Agazie et al, The NANOGrav 15 yr Data Set: Constraints on Supermassive Black Hole Binaries from the Gravitational-wave Background, The Astrophysical Journal Letters (2023). DOI: 10.3847/2041-8213/ace18b


\bibitem{ZN-PBH}
Ya.B. Zeldovich, I.D. Novikov
{ "The Hypothesis of Cores Retarded During Expansion and the Hot Cosmological Model", 
Astronomicheskij Zhurnal, 43 (1966) 758,
 Soviet Astronomy, AJ.10(4):602–603;(1967)}.

\bibitem{SH-PBH}
S.~Hawking,
"Gravitationally collapsed objects of very low mass",
Mon.\ Not.\ Roy.\ Astron.\ Soc.\  {\bf 152}, 75 (1971).

\bibitem{Carr-SH}
B.~J.~Carr and S.~W.~Hawking,
	"Black holes in the early Universe,''
	Mon.\ Not.\ Roy.\ Astron.\ Soc.\  {\bf 168}, 399 (1974).

\bibitem{carr-infl}
B.J. Carr, J.H. Hilbert, J.E. Lidsey, "Black hole relics and inflation: Limits on blue perturbation spectra'',
Phys.Rev.D  50 (1994) 4853, astro-ph/9405027.

\bibitem{INN}
P. Ivanov, P. Naselsky, I. Novikov, Inflation and primordial black holes as dark matter,
PRD 50 (1994) 7173.

\bibitem{AD-KP}
A. Dolgov, K. Postnov, Why the mean mass of primordial black hole distribution is close to 10 $M_\odot$,
JCAP 07 (2020) 063,  e-Print: 2004.11669 [astro-ph.CO].

\bibitem{arefyeva}
I. Ya. Aref'eva, A. Hajilou, K. Rannu, P.l Slepov
"Magnetic catalysis in holographic model with two types of anisotropy for heavy quarks",
Eur.Phys.J.C 83 (2023) 12, 1143, e-Print: 2305.06345 [hep-th].


\bibitem{SH-BH}
S. Hawking, "Gravitationally collapsed objects of very low mass",
Mon. Not. R. astr. Soc. (1971) 152, 75

\bibitem{chaplin}
G.F. Chapline, 
"Cosmological effects of primordial black holes".
Nature, 253, 251 (1975).



\bibitem{BH-limits}
B. Carr, F. Kuhnel
"Primordial Black Holes as Dark Matter: Recent Developments",
arXiv:2006.02838.

\bibitem{boehm}
C. Boehm et al,
"Eliminating the LIGO bounds on primordial black hole dark matter",
JCAP 03 (2021) 078,  arXiv:2008.10743. 

\bibitem{frampton}
C. Corian{\`o}, P.H. Frampton, 
''Does CMB Distortion Disfavour Intermediate Mass Dark Matter?'',
arXiv:2012.13821 [astro-ph.GA]

 \bibitem{rubin}
S.G. Rubin, at al 
''The Formation of Primary Galactic Nuclei during Phase Transitions in the Early Universe'',  
Soviet Journal of Experimental and Theoretical Physics. 2001, V. 92, no. 6. 921;
arXiv:hep-ph/0106187.  

\bibitem{eroshenko}
Y. Eroshenko, V. Stasenko, 
'Gravitational waves from the merger of two primordial black hole clusters'',
Symmetry 15 (2023) 3, 637, 
arXiv:2302.05167.
 
\bibitem{Bellomo}
N. Bellomo, J.L. Bernal,  A. Raccanelli, L. Verde, 
"Primordial Black Holes as Dark Matter: Converting Constraints from Monochromatic to 
Extended Mass Distributions"
JCAP 01 (2018) 004.

\bibitem{KF}
F.~K$\ddot u$hnel, K. Freese,
"Constraints on Primordial Black Holes with {\bf Extended} Mass Functions",
Phys. Rev. D 95 (2017) 8, 083508, e-Print: 1701.07223.

\bibitem{BDPP}
S.Blinnkov, A.Dolgov,  N.Porayko, K.Postnov, "Solving puzzles of GW150914 by primordial black holes,"  
JCAP 1611 (2016), 036, e-Print: 1611.00541.

\bibitem{Post-Mit}
K. Postnov, N. Mitichkin, 
Spins of primordial binary black holes before coalescence, 
JCAP 1906 (2019) no.06, 044,  arXiv 1904.00570 [astro-ph.HE]

\bibitem{PKM}
K. Postnov, A. Kuranov, N. Mitichkin, 
"Spins of black holes in coalescing compact binaries",
Physics-Uspekhi vol. 62, No. 11, (2019),  arXiv:1907.04218 .

\bibitem{dkmppss}
A.D. Dolgov, A.G. Kuranov, N.A. Mitichkin, S. Porey, K.A. Postnov, O.S.~Sazhina,  I.V. Simkine,
''On mass distribution of coalescing black holes'',
JCAP 12 (2020) 017,  e-Print: 2005.00892. 

\bibitem{Raidal}
M. Raidal et al, 
''Formation and Evolution of Primordial Black Hole Binaries in the Early Universe,''
JCAP. V. 2019, no. 2. P. 018. arXiv:1812.01930

\bibitem{Liu}
L. Liu, et al 
''Constraining the Merger History of Primordial-Black-Hole Binaries from GWTC-3'',
arXiv:2210.16094.

\bibitem{KAP-Mit} 
K. Postnov, N. Mitichkin, '''On the primordial binary black hole mergings in LVK data'',  
e-Print: 2302.06981 [astro-ph.CO].

\bibitem{DVN}
A.D. Dolgov, V.A. Novikov, M.I. Vysotsky, ''How to see an antistar''
JETP Lett. 98 (2013) 519, e-Print: 1309.2746 

\bibitem{shuster}
A. Schuster, Nature, 58 (1898) 367. 
''Potential Matter. Holiday Dream.''
Nature, 58 (1898) 367. 

\bibitem{sakharov}
A. D. Sakharov, 
Violation of CP Invariance, C Asymmetry, and Baryon Asymmetry of the Universe
Pisma v Zhurnal Eksperimentalnoi i Teoreticheskoi Fiziki 5, 32 (1967); Soviet Journal of Experimental
and Theoretical Physics Letters 5, 24 (1967).

\bibitem{AD-CP}
A.D. Dolgov,
''CP violation in cosmology'',
Contribution to: 163rd Course of International School of Physics ''Enrico Fermi'', 407-438, 
e-Print: hep-ph/0511213 [hep-ph].

\bibitem{stecker-1}
F. W. Stecker,  D. L. Morgan, Jr., J. Bredekamp, 
'' Possible Evidence for the Existence of
Antimatter on a Cosmological Scale in the Universe'', Phys. Rev. Letters 27, 1469 (1971)

\bibitem{konstan-1}
B.P. Konstantinov, et al 
 Cosmic Research, 4, 66 (1968).

\bibitem{konstan-2}
B.P. Konstantinov, et al
Bulletin of the Academy of Sciences of
the USSR. physical series, {\bf 33}, No.11, 1820 (1969)

\bibitem{stecker-2}
F. W. Stecker, "Grand Unification and possible matter-antimatter domain structure in the
universe". Tenth Texas Symposium on Relativistic Astrophysics, p. 69 (1981).

\bibitem{C-AdR-G}
A.G. Cohen, A. De Rujula, S.L. Glashow,
"A Matter - antimatter universe?"
Astrophys.J. 495 (1998) 539, e-Print: astro-ph/9707087 [astro-ph] 

\bibitem{GS-1}
G. Steigman, ”Observational tests of antimatter cosmologies”, 
Ann. Rev. Astron. Astrophys. 14, 339 (1976).

\bibitem{GS-2}
G. Steigman, ”When Clusters Collide: Constraints On Antimatter On The Largest Scales”, JCAP 10, 001
(2008); e-Print: 0808.1122 [astro-ph].

\bibitem{stecker-Blois}
F. W. Stecker, ”The Matter-Antimatter Asymmetry of the Universe (keynote address for XIVth
Rencontres de Blois)” arXiv:hep-ph/0207323.

\bibitem{dolgov-Blois}
A.D. Dolgov, ”Cosmological matter antimatter asymmetry and antimatter in the universe”,
keynote lecture at 14th Rencontres de Blois on Matter - Anti-matter Asymmetry, e-Print: hep- ph/0211260.

\bibitem{ballmoos-bound}
P. von Ballmoos, ”Antimatter in the Universe : Constraints from Gamma-Ray Astronomy”, 
 Hyperfine Interact. 228 (2014) 1-3, 91-100, Contribution to: LEAP2013, 91-100, e-Print: 1401.7258 

\bibitem{bd}
C.Bambi, A.D. Dolgov,
{ "Antimatter in the Milky Way",}
Nucl.Phys.B 784 (2007) 132-150, astro-ph/0702350.

\bibitem{db}
A.D. Dolgov, S.I. Blinnikov,
{ "Stars and Black Holes from the very Early Universe",}
Phys.Rev.D 89 (2014) 2, 021301, arXive 1309.3395.

\bibitem{bdp}
S.I.Blinnikov, A.D. Dolgov, K.A.Postnov,
{"Antimatter and antistars in the universe and in the Galaxy",}
Phys.Rev.D 92 (2015) 023516, arXive  1409.5736.

\bibitem{anti-e1}
J. Kn\"{o}dlseder \textit{et al.},
\textit{The all-sky distribution of 511 keV electron-positron annihilation emission},
Astron. Astrophys. \textbf{441} (2005) 513, arXiv:astro-ph/0506026.

\bibitem{anti-e2}
P. Jean \textit{et al.},
\textit{Spectral analysis of the Galactic $e^+ e^-$ annihilation emission},
Astron. Astrophys. \textbf{445} (2006) 579, arXiv:astro-ph/0509298.

\bibitem{anti-e3}
G. Weidenspointner \textit{et al.},
\textit{The sky distribution of positronium annihilation continuum emission measured with SPI/INTEGRAL},
Astron. Astrophys. \textbf{450} (2006) 1013, arXiv:astro-ph/0601673.

\bibitem{gr-ann}
I.\,F. Mirabel, 
\textit{The Great Annihilator in the Central Region of the Galaxy},
The Messenger \textbf{70} (1992) 51.

\bibitem{Ein-observ}
P. Hertz, J.\,E. Grindlay,
\textit{The Einstein galactic plane survey: statistical analysis of the complete X-ray sample},
Astrophys. J. \textbf{278} (1984) 137.

\bibitem{granat}
R. Sunyaev \textit{et al.},
\textit{Two hard X-ray sources in 100 square degrees around the Galactic Center},
Astron. Astrophys. \textbf{247} (1991) L29.

\bibitem{AMS-19}
M. Aguilar \textit{et al.} (AMS Collaboration),
\textit{Towards Understanding the Origin of Cosmic-Ray Positrons},
Phys. Rev. Lett. \textbf{122} (2019) 041102.

\bibitem{kachel}
M. Kachelriess, A. Neronov, D.V. Semikoz, 
''Signatures of a two million year old supernova in the spectra of cosmic ray protons, antiprotons and positrons'',
Phys.Rev.Lett. 115 (2015) 18, 181103 • e-Print: 1504.06472 [astro-ph.HE].


\bibitem{BDP}
C. Bambi, A.\,D. Dolgov, A.\,A. Petrov,
\textit{Black holes as antimatter factories}, 
JCAP \textbf{09} (2009) 013, arXiv:0806.3440 [astro-ph].

\bibitem{AD-AR}
A.D. Dolgov, A.S. Rudenko,
''Conversion of protons to positrons by a black hole'',
 2308.01689 [hep-ph].

\bibitem{anti-nuc-AMS-1}
S. Ting, 
\textit{A Brief Summary of Ten Years of New and Unexpected Results from the Alpha Magnetic Spectrometer on the International Space Station}, 
proceedings of 44th COSPAR Scientific Assembly (16-24 July 2022), vol. 44, p. 3071.

\bibitem{anti-nuc-AMS-2}
V. Choutko, 
\textit{Cosmic Heavy Anti-Matter}, 
proceedings of 44th COSPAR Scientific Assembly (16-24 July 2022), vol. 44, p. 2083


\bibitem{anti-stars}
S.~Dupourqu{\'e}, L.~{Tibaldo} and P.~{von Ballmoos}, ''Constraints on
  the antistar fraction in the Solar System neighborhood  from the 10-year Fermi
  Large Area Telescope gamma-ray source catalog'',
    Phys Rev D.103.083016 103 (2021) 083016 

\bibitem{bbbdp}
A.E. Bondar, S.I. Blinnikov, A.M. Bykov, A.D. Dolgov, K.A. Postnov,
''X-ray signature of antistars in the Galaxy''
JCAP 03 (2022) 03, 009, JCAP 03 (2022) 009, 
e-Print: 2109.12699.

\bibitem{anti-BBN}
A. Arbey, J. Auffinger, J. Silk
"Stellar signatures of inhomogeneous big bang nucleosynthesis
Phys.Rev.D 102 (2020) 2, 023503, e-Print: 2006.02446 [astro-ph.CO]


\bibitem{anti-bbn}
V. Poulin, P. Salati, I. Cholis, M. Kamionkowski, J. Silk
"Where do the AMS-02 antihelium events come from?"
Phys. Rev. D 99, 023016 (2019),
arXiv:1808.08961 [astro-ph.HE].


\bibitem{bpbbd}
A.M. Bykov, K.A. Postnov, A.E. Bondar, S.I. Blinnikov, A.D. Dolgov, 
Antistars as possible sources of
antihelium cosmic rays, 
JCAP 08 (2023) 027, e-Print: 2304.04623 [astro-ph.HE]

\bibitem{choutko}
A. Choutko,
AMS-02 Collaboration, “AMS Days at La Palma, La Palma, Canary Islands, Spain,” (2018).

\bibitem{ting-CERN}
S. Ting, Latest Results from the AMS Experiment on the International Space Station. Colloquium at CERN, May, 2018.


\bibitem{cosm-anti-nuc}
R. Duperray, B. Baret, D. Maurin, et al,
"Flux of light antimatter nuclei near Earth, induced by cosmic rays in the Galaxy and in the atmosphere",
 Phys. Rev.D 71 (2005) 083013, e-Print: astro-ph/0503544.

\bibitem{antistars}
S.~Dupourqu{\'e}, L.~{Tibaldo} and P.~{von Ballmoos}, 
'"Constraints on the antistar fraction in the Solar System neighborhood 
 from the 10-year Fermi Large Area Telescope gamma-ray source catalog", Phys. Rev. D.103.083016 103 (2021) 083016,  e-Print: 2103.10073 [astro-ph.HE].

\bibitem{AD}
I. Affleck, M. Dine, ''A New Mechanism for Baryogenesis'',
Nucl. Phys. B 249 (1985) 361-380.

\bibitem{vil}
 A. Vilenkin and L.H. Ford, Gravitational effects upon cosmological phase transitions,
Phys.Rev. D26 (1982) 1231-1241.

\bibitem{adl}
A.D. Linde,'' Scalar field fluctuations in the expanding universe and the new inflationary universe scenario'', 
Phys.Lett. B116 (1982) 335-339;

\bibitem{aas}
A.A. Starobinski, ''Dynamics of phase transitions in the new inflationary universe scenario and
generation of perturbations'', Phys. Lett, B117 (1982) 175-178;

\bibitem{ad-dp}
A. Dolgov, D.N. Pelliccia
''Scalar field instability in de Sitter space-time'',
Nucl.Phys.B 734 (2006) 208-219 • e-Print: hep-th/0502197 [hep-th].

\end{thebibliography}
\end{document}